\documentclass[twocolumn,aps,unsortedaddress,superscriptaddress,prd,floatfix,showpacs,nofootinbib]{revtex4-1} % ,linenumbers
\usepackage{epsfig}
\usepackage[utf8]{inputenc}
\usepackage{float}
\usepackage{graphicx}
\usepackage{amsmath}
\usepackage{verbatim}
\usepackage[dvipsnames]{xcolor}
\usepackage{hyperref}
\usepackage{footmisc}
\begin{document}

\title{Measurement of flavor asymmetry of light-quark sea in the proton with Drell-Yan dimuon production in $p+p$ and $p+d$ collisions at 120 GeV}

\author{J. Dove}
\affiliation{Department of Physics, University of Illinois at Urbana-Champaign,
Urbana, Illinois 61801, USA}

\author{B.~Kerns}
\affiliation{Department of Physics, University of Illinois at Urbana-Champaign,
Urbana, Illinois 61801, USA}

\author{C.~Leung}
\affiliation{Department of Physics, University of Illinois at Urbana-Champaign, Urbana, Illinois 61801, USA}

\author{R.~E.~McClellan}
\altaffiliation[Now at ]{Pensacola State College, Pensacola, FL 32504}
\affiliation{Department of Physics, University of Illinois at Urbana-Champaign, 
Urbana, Illinois 61801, USA}

\author{S.~Miyasaka}
\affiliation{Department of Physics, Tokyo Institute of Technology, Meguro-ku, Tokyo 152-8550, Japan}

\author{D.~H.~Morton}
\affiliation{Randall Laboratory of Physics, University of Michigan, Ann Arbor,
Michigan 48109, USA}

\author{K.~Nagai}
\affiliation{Department of Physics, Tokyo Institute of Technology, Meguro-ku, Tokyo 152-8550, Japan}
\affiliation{Institute of Physics, Academia Sinica, Taipei, 11529, Taiwan}
\affiliation{Physics Division, Los Alamos National Laboratory, Los Alamos, New Mexico 97545, USA}

\author{S.~Prasad}
\affiliation{Department of Physics, University of Illinois at Urbana-Champaign,
Urbana, Illinois 61801, USA}
\affiliation{Physics Division, Argonne National Laboratory, Lemont, Illinois 60439, USA}

\author{F.~Sanftl}
\affiliation{Department of Physics, Tokyo Institute of Technology, Meguro-ku,
Tokyo 152-8550, Japan}

\author{M.~B.~C.~Scott}
\affiliation{Randall Laboratory of Physics, University of Michigan, Ann Arbor,Michigan 48109, USA}
\affiliation{Physics Division, Argonne National Laboratory, Lemont, Illinois 60439, USA}

\author{A.~S.~Tadepalli}
\altaffiliation[Now at ]{Thomas Jefferson National Accelerator Facility, Newport News, Virginia 23606, USA}
\affiliation{Department of Physics and Astronomy, Rutgers, The State University of New Jersey, Piscataway, New Jersey 08854, USA}

\author{C.~A.~Aidala}
\affiliation{Randall Laboratory of Physics, University of Michigan, Ann Arbor, Michigan 48109, USA}
\affiliation{Physics Division, Los Alamos National Laboratory, Los Alamos, New Mexico 97545, USA}

\author{J.~ Arrington}
\altaffiliation[Now at ]{Lawrence Berkeley National Laboratory, Berkeley, California, 94720 USA}
\affiliation{Physics Division, Argonne National Laboratory, Lemont, Illinois 60439, USA}

\author{C.~Ayuso}
\affiliation{Randall Laboratory of Physics, University of Michigan, Ann Arbor,
Michigan 48109, USA}

\author{C.~T.~Barker}
\affiliation{Department of Engineering and Physics, Abilene Christian University, Abilene, Texas 79699, USA}

\author{C.~N.~Brown}
\affiliation{Fermi National Accelerator Laboratory, Batavia, Illinois 60510, USA}

\author{T.~H.~Chang}
\affiliation{Institute of Physics, Academia Sinica, Taipei, 11529, Taiwan}

\author{W.~C.~Chang}
\affiliation{Institute of Physics, Academia Sinica, Taipei, 11529, Taiwan}

\author{A.~Chen}
\affiliation{Department of Physics, University of Illinois at Urbana-Champaign, Urbana, Illinois 61801, USA}
\affiliation{Institute of Physics, Academia Sinica, Taipei, 11529, Taiwan}
\affiliation{Randall Laboratory of Physics, University of Michigan, Ann Arbor, Michigan 48109, USA}

\author{D.~C.~Christian}
\affiliation{Fermi National Accelerator Laboratory, Batavia, Illinois 60510, USA}

\author{B.~P.~Dannowitz}
\affiliation{Department of Physics, University of Illinois at Urbana-Champaign, Urbana, Illinois 61801, USA}

\author{M.~Daugherity}
\affiliation{Department of Engineering and Physics, Abilene Christian University, Abilene, Texas 79699, USA}

\author{M.~Diefenthaler}
\affiliation{Department of Physics, University of Illinois at Urbana-Champaign, Urbana, Illinois 61801, USA}

\author{L.~El Fassi}
\affiliation{Department of Physics and Astronomy, Mississippi State University, Mississippi State, Mississippi 39762, USA }
\affiliation{Department of Physics and Astronomy, Rutgers, The State University of New Jersey, Piscataway, New Jersey 08854, USA}

\author{D.~F.~Geesaman}
\affiliation{Physics Division, Argonne National Laboratory, Lemont, Illinois 60439, USA}

\author{R.~Gilman}
\affiliation{Department of Physics and Astronomy, Rutgers, The State University of New Jersey, Piscataway, New Jersey 08854, USA}

\author{Y.~Goto}
\affiliation{RIKEN Nishina Center for Accelerator-Based Science, Wako, Saitama 351-0198, Japan}

\author{L.~Guo}
\altaffiliation[Now at ]{Florida International University, Miami, Florida, 33199, USA}
\affiliation{Physics Division, Los Alamos National Laboratory, Los Alamos, New Mexico 97545, USA}

\author{R.~Guo}
\affiliation{Department of Physics, National Kaohsiung Normal University, Kaohsiung City 80201, Taiwan}

\author{T.~J.~Hague}
\altaffiliation[Now at ]{Lawrence Berkeley National Laboratory, Berkeley, California, 94720 USA}
\affiliation{Department of Engineering and Physics, Abilene Christian University, Abilene, Texas 79699, USA}

\author{R.~J.~Holt}
\altaffiliation[Now at ]{Kellogg Radiation Laboratory, California Institute of Technology, Pasadena, California 91125, USA}
\affiliation{Physics Division, Argonne National Laboratory, Lemont, Illinois 60439, USA}

\author{D.~Isenhower}
\affiliation{Department of Engineering and Physics, Abilene Christian University, Abilene, Texas 79699, USA}

\author{E.~R.~Kinney}
\affiliation{Department of Physics, University of Colorado, Boulder,
Colorado 80309, USA}

\author{N.~D.~Kitts}
\affiliation{Department of Engineering and Physics, Abilene Christian University, Abilene, Texas 79699, USA}

\author{A.~Klein}
\affiliation{Physics Division, Los Alamos National Laboratory, Los Alamos, New Mexico 97545, USA}

\author{D.~W.~Kleinjan}
\affiliation{Physics Division, Los Alamos National Laboratory, Los Alamos, New Mexico 97545, USA}

\author{Y.~Kudo}
\affiliation{Department of Physics, Yamagata University, Yamagata City, Yamagata 990-8560, Japan}

\author{P.-J.~Lin}
\affiliation{Department of Physics, University of Colorado, Boulder, Colorado 80309, USA}
\affiliation{Institute of Physics, Academia Sinica, Taipei, 11529, Taiwan}

\author{K.~Liu}
\affiliation{Physics Division, Los Alamos National Laboratory, Los Alamos, New Mexico 97545, USA}

\author{M.~X.~Liu}
\affiliation{Physics Division, Los Alamos National Laboratory, Los Alamos,
New Mexico 97545, USA}

\author{W.~Lorenzon}
\affiliation{Randall Laboratory of Physics, University of Michigan, Ann Arbor,
Michigan 48109, USA}

\author{N.~C.~R.~Makins}
\affiliation{Department of Physics, University of Illinois at Urbana-Champaign, Urbana, Illinois 61801, USA}

\author{M.~Mesquita de Medeiros}
\affiliation{Physics Division, Argonne National Laboratory, Lemont,
Illinois 60439, USA}

\author{P.~L.~McGaughey}
\affiliation{Physics Division, Los Alamos National Laboratory, Los Alamos,
New Mexico 97545, USA}

\author{Y.~Miyachi}
\affiliation{Department of Physics, Yamagata University, Yamagata City,
Yamagata 990-8560, Japan}

\author{I.~Mooney}
\altaffiliation[Now at ]{Wayne State University, Detroit, MI 48202, USA}
\affiliation{Randall Laboratory of Physics, University of Michigan, Ann Arbor, Michigan 48109, USA}

\author{K.~Nakahara}
\altaffiliation[Now at ]{Stanford Linear Accelerator Center, Menlo Park, CA 94025, USA}
\affiliation{Department of Physics, University of Maryland, College Park, Maryland 20742, USA}

\author{K.~Nakano}
\affiliation{University of Virginia, Charlottesville, Virginia 22904, USA}
\affiliation{Department of Physics, Tokyo Institute of Technology, Meguro-ku, Tokyo 152-8550, Japan}
\affiliation{RIKEN Nishina Center for Accelerator-Based Science, Wako, Saitama 351-0198, Japan}

\author{S.~Nara}
\affiliation{Department of Physics, Yamagata University, Yamagata City, Yamagata 990-8560, Japan}

\author{J.~C.~Peng}
\affiliation{Department of Physics, University of Illinois at Urbana-Champaign,
Urbana, Illinois 61801, USA}

\author{A.~J.~Puckett}
\altaffiliation[Now at ]{University of Connecticut, Storrs, CT 06269, USA}
\affiliation{Physics Division, Los Alamos National Laboratory, Los Alamos, New Mexico 97545, USA}

\author{B.~J.~Ramson}
\affiliation{Randall Laboratory of Physics, University of Michigan, Ann Arbor, Michigan 48109, USA}
\affiliation{Fermi National Accelerator Laboratory, Batavia, Illinois 60510, USA}

\author{P.~E.~Reimer}
\affiliation{Physics Division, Argonne National Laboratory, Lemont, Illinois 60439, USA}

\author{J.~G.~Rubin}
\affiliation{Randall Laboratory of Physics, University of Michigan, Ann Arbor, Michigan 48109, USA}
\affiliation{Physics Division, Argonne National Laboratory, Lemont, Illinois 60439, USA}

\author{S.~Sawada}
\affiliation{Institute of Particle and Nuclear Studies, KEK, High Energy Accelerator Research Organization, Tsukuba, Ibaraki 305-0801, Japan}

\author{T.~Sawada}
\altaffiliation[Now at ]{Institute for Cosmic Ray Research, KAGRA Observatory, The University of Tokyo, Hida, Gifu 506-1205, Japan}
\affiliation{Randall Laboratory of Physics, University of Michigan, Ann Arbor, Michigan 48109, USA}

\author{T.-A.~Shibata}
\altaffiliation[Now at ]{Nihon University, College of Science and Technology, Chiyoda-ku, Tokyo 101-8308, Japan}
\affiliation{Department of Physics, Tokyo Institute of Technology, Meguro-ku, Tokyo 152-8550, Japan}
\affiliation{RIKEN Nishina Center for Accelerator-Based Science, Wako, Saitama 351-0198, Japan}

\author{S.~H.~Shiu}
\affiliation{Institute of Physics, Academia Sinica, Taipei, 11529, Taiwan}

\author{D.~Su}
\affiliation{Institute of Physics, Academia Sinica, Taipei, 11529, Taiwan}

\author{M.~Teo}
\affiliation{Department of Physics, University of Illinois at Urbana-Champaign, Urbana, Illinois 61801, USA}

\author{B.~G Tice}
\affiliation{Physics Division, Argonne National Laboratory, Lemont, Illinois 60439, USA}

\author{R.~S.~Towell}
\affiliation{Department of Engineering and Physics, Abilene Christian University, Abilene, Texas 79699, USA}

\author{S.~Uemura}
\altaffiliation[Now at ]{Fermi National Accelerator Laboratory, Batavia, Illinois 60510, USA}
\affiliation{Department of Engineering and Physics, Abilene Christian University, Abilene, Texas 79699, USA}

\author{T.~S.~Watson}
\affiliation{Department of Engineering and Physics, Abilene Christian University, Abilene, Texas 79699, USA}

\author{S.~G.~Wang}
\altaffiliation[Now at ]{APS, Argonne National Laboratory, Lemont, Illinois 60439, USA}
\affiliation{Institute of Physics, Academia Sinica, Taipei, 11529, Taiwan}
\affiliation{Department of Physics, National Kaohsiung Normal University, Kaohsiung City 80201, Taiwan}

\author{A.~B.~Wickes}
\affiliation{Physics Division, Los Alamos National Laboratory, Los Alamos,
New Mexico 97545, USA}

\author{J.~Wu}
\affiliation{Fermi National Accelerator Laboratory, Batavia, Illinois 60510, USA}

\author{Z.~Xi}
\affiliation{Department of Engineering and Physics, Abilene Christian University, Abilene, Texas 79699, USA}

\author{Z.~Ye}
\altaffiliation[Now at ]{Department of Physics, Tsinghua University, Beijing 100084 China}
\affiliation{Physics Division, Argonne National Laboratory, Lemont, Illinois 60439, USA}

\collaboration{FNAL E906/SeaQuest Collaboration}
%\noaffiliation

\date{\today}


\begin{abstract}
Evidence for a flavor asymmetry between the $\bar u$ and $\bar d$ quark distributions in the proton has been found in deep-inelastic scattering and Drell-Yan experiments. The pronounced dependence of this flavor asymmetry on $x$ (fraction of nucleon momentum carried by partons) observed in the Fermilab E866 Drell-Yan experiment suggested a drop of the $\bar d\left(x\right) / \bar u\left(x\right)$ ratio in the $x > 0.15$ region. We report results from the SeaQuest Fermilab E906 experiment with improved statistical precision for $\bar d\left(x\right) / \bar u\left(x\right)$ in the large $x$ region up to $x=0.45$ using the 120 GeV proton beam.  Two different methods for extracting the Drell-Yan cross section ratios, $\sigma^{pd} /2 \sigma^{pp}$, from the SeaQuest data give consistent results. The $\bar{d}\left(x\right) / \bar{u}\left(x\right)$ ratios and the $\bar d\left(x\right) - \bar u\left(x\right)$ differences are deduced from these cross section ratios for $0.13 < x < 0.45$. The SeaQuest and E866/NuSea $\bar{d}\left(x\right) / \bar{u}\left(x\right)$ ratios are in good agreement for the $x\lesssim 0.25$ region. The new SeaQuest data, however, show that $\bar d\left(x\right)$ continues to be greater than $\bar u\left(x\right)$ up to the highest $x$ value ($x = 0.45$).  The new results on $\bar{d}\left(x\right) / \bar{u}\left(x\right)$ and $\bar{d}\left(x\right) - \bar{u}\left(x\right)$ are compared with various parton distribution functions and theoretical calculations. 
\end{abstract} 
\maketitle
\clearpage

\section{Introduction}

The first direct experimental evidence for the existence of sea-quark distributions in the nucleon was the observation of the rise of the nucleon structure functions at small momentum fraction (Bjorken-$x$) in deep-inelastic scattering (DIS) experiments~\cite{Friedman72}. The advent of Quantum Chromodynamics (QCD) as the theory of strong interactions~\cite{Gross73,Politzer73} offered a natural explanation for these sea quarks as originating from the gluon splitting into a quark and antiquark pair. As a result of the large coupling between gluons and quarks in the strong interaction, the sea quarks feature prominently in the description of the internal structure of the proton. Together with the valence-quark and gluon distributions, the sea-quark distributions in the proton have been explored in numerous experiments using DIS, the Drell-Yan process~\cite{Drell70,Drell-YanAnnals} and other hard processes~\cite{Chang}.

While it was well established that the valence quark distributions in the proton are different for up and down quarks, it was assumed that the sea quark distribution is flavor symmetric, i.e., $\bar u\left(x\right) = \bar d\left(x\right)$. This assumption did not arise from fundamental symmetry principles. Rather, it was based on the expectation that the $g \to q \bar q$ QCD process should yield similar $\bar u$ and $\bar d$ sea-quark content because the coupling of strong interaction is flavor independent and  the phase space of the $g \to q \bar q$ process is nearly the same for $\bar u$ and $\bar d$ as the masses of these light quarks are small and comparable. It was a major surprise when the NMC  Collaboration~\cite{NMC91,NMC2} reported a precise DIS measurement showing the violation of the Gottfried Sum Rule~\cite{GSR,FieldFeynman},  suggesting an asymmetry between the $\bar u$ and the $\bar d$  distributions in the proton.

\begin{figure}
    \centering
    \includegraphics[width=\columnwidth]{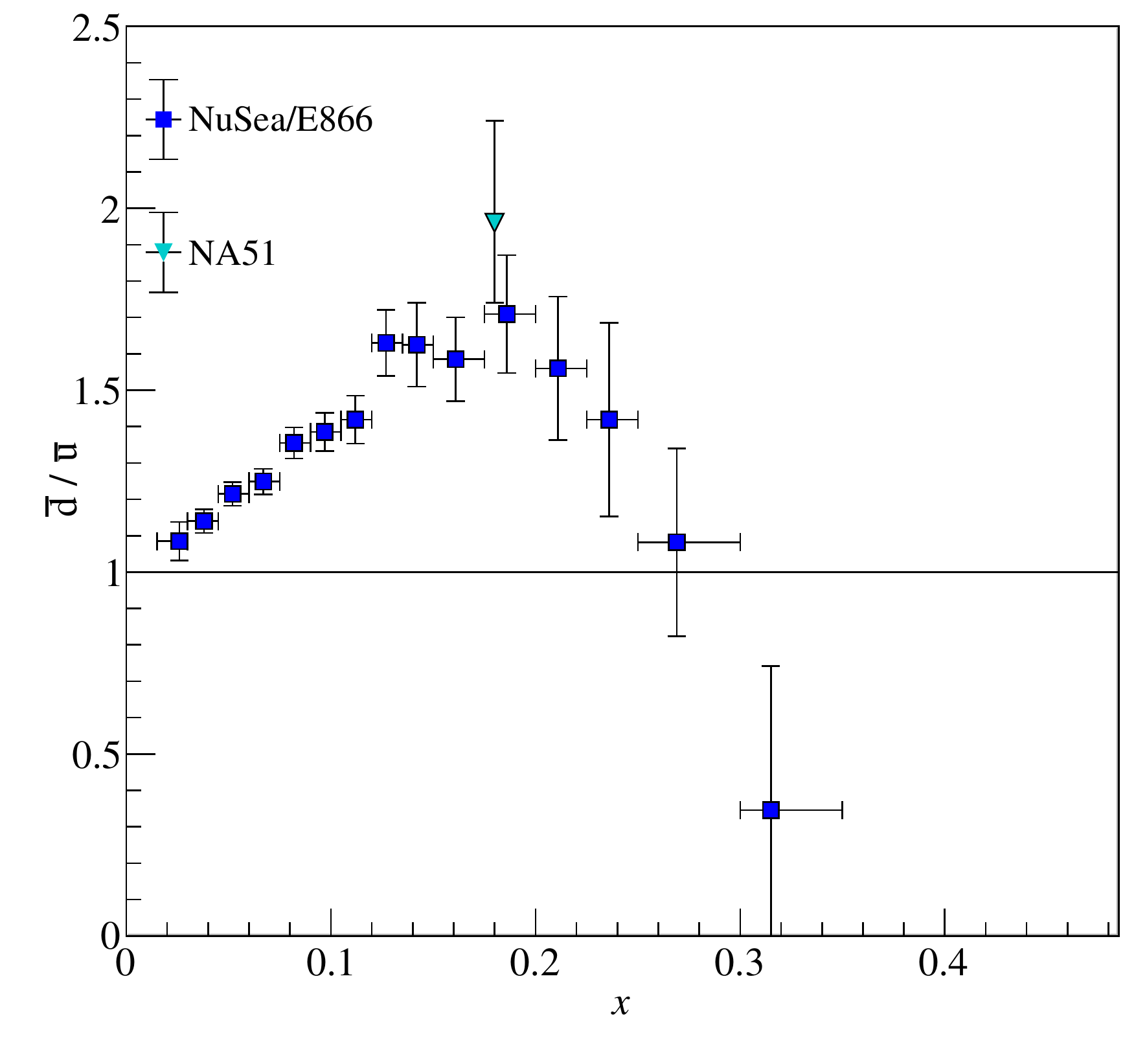}
    \caption{The $\bar{d}/\bar{u}$ ratio versus $x$ extracted from NA51~\cite{NA51} and E866~\cite{PhysRevLett.80.3715,Towell:2001nh,Peng:1998pa}. \label{fig:early_data}}
\end{figure}

To verify the surprising result from the NMC on the asymmetry between $\bar u$ and $\bar d$ distributions in the proton, an independent experimental approach, namely the Drell-Yan process with a proton beam, was suggested~\cite{Ellis}. From a comparison between the $\mu^+\mu^-$ (dimuon) production cross sections in proton-proton ($p+p$) and proton-deuterium ($p+d$) collisions, it was shown by the NA51~\cite{NA51} and the E866~\cite{PhysRevLett.80.3715,Towell:2001nh,Peng:1998pa} Collaborations that $\bar u \left(x\right)$ is indeed different from $\bar d\left(x\right)$, as plotted in Fig.~\ref{fig:early_data}. \footnote{It is generally clear from context when the symbol ``$d$'' denotes the deuteron and when it denotes a(n) (anti)quark distribution.  When extra clarity is needed, the quark distributions will be written as a function of $x$, i.e., $d(x)$.} By covering a broad kinematic range, the E866 experiment revealed an intriguing $x$ dependence of the $\bar d\left(x\right) / \bar u\left(x\right)$ asymmetry. At the lowest $x$ values, the $\bar u\left(x\right)$ and $\bar d\left(x\right)$ are comparable. As $x$ increases, $\bar d\left(x\right) / \bar u\left(x\right)$ rises monotonically, reaching a maximal value of about 1.75 at $x\approx 0.15$. At even higher $x$ values, the E866 data indicate that $\bar d\left(x\right) / \bar u\left(x\right)$ starts to drop, becoming less than unity at the highest $x$ value measured~\cite{PhysRevLett.80.3715, Towell:2001nh, Peng:1998pa}.  The flavor asymmetry between $\bar d(x)$ and $\bar u(x)$ was also observed by the HERMES collaboration using semi-inclusive deep inelastic scattering~\cite{Hermes}.

The results from the NMC, NA51, E866, and HERMES experiments inspired many theoretical attempts to explain the surprising flavor asymmetry of the light-quark sea and its $x$ dependence. Recent reviews on this  subject can be found in Ref.~\cite{Chang,Geesaman:2018ixo}. These data have also been included in global fits that extract the parton distribution functions (PDFs) of the proton. All modern PDFs now allow for an asymmetry between $\bar u\left(x\right)$ and $\bar d\left(x\right)$. Lattice QCD calculations of the $\bar d\left(x\right) - \bar u\left(x\right)$ based on the Large Momentum Effective theory (LaMET)~\cite{Ji17} have also recently become available~\cite{Lin18,Cyprus21}, confirming an asymmetry between $\bar u\left(x\right)$ and $\bar d\left(x\right)$. A recent calculation of proton PDFs using a quark-diquark representation of the proton's Poincar\`e-covariant wave function can also describe the $\bar d / \bar u$ asymmetry~\cite{Roberts}.

Several models, including the meson cloud model~\cite{Thomas83,Henley90,Kumano91,Wally99},  chiral-quark model~\cite{Eichten92,Cheng95,Szczurek96},  statistical model~\cite{Bourrely:2002vg,Bourrely:2005tk}, chiral-quark soliton  model~\cite{Diakonov96,Wakamatsu03}, and instanton  model~\cite{Schaefer98,Dorokhov93} are capable of describing the rise of $\bar d\left(x\right) / \bar u\left(x\right)$ as $x$ increases in the small $x$ region. (See Refs.~\cite{thomas,kumano,garvey,Chang} for reviews.) In most of these models a meson cloud is the source of the flavor asymmetry between $\bar d\left(x\right)$ and $\bar u\left(x\right)$. While the importance of the meson cloud for understanding the nucleon form factors at a relatively low $Q^2$ scale is well established~\cite{Pion}, it is interesting that the meson cloud could also lead to a flavor asymmetry in the partonic structures probed at much higher $Q^2$ scales~\cite{Thomas83,Sullivan}. None of these models can describe the drop of $\bar d\left(x\right) / \bar u\left(x\right)$ below unity at large $x$ indicated by the E866 data, plotted in Fig.~\ref{fig:early_data}. The highest $x$ data points from E866 contain large statistical uncertainties and so a new measurement with improved precision for the large $x$ region was warranted. 

The Drell-Yan cross section for fixed values of $x_1$ and $x_2$, which refer to the beam and target Bjorken-$x$ respectively, is inversely proportional to $s$, the hadron-hadron center-of-mass energy squared. Therefore, proton beams at lower energies are able to probe the large $x$ region, where the measurement is limited by small cross sections. The small $x$ region has already been measured with high precision by E866, which used an 800 GeV proton beam. The SeaQuest (Fermilab E906) experiment, using a new spectrometer~\cite{E906trig2} and a 120 GeV proton beam, finished recording data in 2017. The first result from SeaQuest was reported in a recent article~\cite{Dove21}.

In this paper, a more detailed report on the SeaQuest experiment and the data analysis, together with additional results are presented. In particular, two different methods for determining the $\sigma^{pd} / 2\sigma^{pp}$ Drell-Yan cross section ratios, which lead to consistent results, are discussed. Moreover, the $\bar{d}\left(x\right)/\bar{u}\left(x\right)$ ratio and the $\bar d\left(x\right) - \bar u\left(x\right)$ difference are deduced from these data and compared with calculations.

After this introduction, the Drell-Yan process and the kinematical variables are briefly summarized in Sec.~\ref{sec:drell-yan}. The SeaQuest experiment is described in Sec.~\ref{sec:apparatus}. Section~\ref{sec:dataAnalysis} presents the data analysis. Results of the Drell-Yan cross section ratios are presented in Sec.~\ref{sec:crossRatio}. Section~\ref{sec:extraction} discusses the extraction of the $\bar d\left(x\right) / \bar u\left(x\right)$ and $\bar d\left(x\right) - \bar u\left(x\right)$ from the measured cross section ratios and compares these these quantities with current PDFs and some theoretical models. We conclude in Sec.~\ref{sec:conclusions}.

\section{The Drell-Yan process \label{sec:drell-yan}}

The SeaQuest experiment detects $\mu^+ \mu^-$ pairs (dimuons) produced in the interaction of a proton beam and various target nuclei. The production of massive $\mu^+ \mu^-$ pairs was described by Drell and Yan~\cite{Drell70} as the annihilation of a quark-antiquark pair into a virtual photon that subsequently decays into a pair of leptons. In the Drell-Yan process, the differential cross section at leading order (LO) is given by
\begin{eqnarray}
\lefteqn{\frac{d^2\sigma}{dx_1dx_2}=\frac{4\pi \alpha^2}{9x_1x_2s} \times} 
\label{eq:DYCross} \\ 
&&\sum_{i\in u,d,s,\dots} e_i^2 \left[q_i^A\left(x_1\right) \bar q_i^B\left(x_2\right) + \bar q_i^A\left(x_1\right)
q_i^B\left(x_2\right)\right], \nonumber 
\end{eqnarray}
where $\alpha$ is the fine-structure constant, $e_i$ is the charge of a quark with flavor $i$, and $q_i^{A,B}\left(x_{1,2}\right)$ are the quark distribution functions in hadrons $A$ and $B$ for quarks carrying a momentum fraction $x_1$ and $x_2$, respectively. 
An analogous notation is used for antiquark distribution functions $\bar q_i^{A,B}\left(x_{1,2}\right)$.

The experiment measures the momenta of the $\mu^+$ and $\mu^-$, $p_{\mu^+}$ and $p_{\mu^-}$.  
From these, the four-momentum $Q$ of virtual photon from the quark-antiquark annihilation is determined, $Q = p_{\mu^+} + p_{\mu^-}$. The variables $x_1$ and $x_2$ of the quark-antiquark pair are then
\begin{eqnarray}
x_1 = \frac{P_2 \cdot Q}{P_2 \cdot P} \mbox{~and~} x_2 = 
\frac{P_1 \cdot Q}{P_1 \cdot P}, 
\label{def_x1x2}
\end{eqnarray}
where $P_1$ and $P_2$ are the four-momenta of the projectile and target hadron, respectively, and $P$ is the sum of $P_1$ and $P_2$, $P=P_1+P_2$. The invariant mass-squared of the dimuon, $M^2 = Q^2$, is related to $x_1$, $x_2$, $s$, and $P_T$ by
\begin{eqnarray}
M^2=x_1 x_2 s - P_T^2,
\label{def_mass}
\end{eqnarray}
where $P_T$ is the transverse momentum of the dimuon. 

The Feynman-$x$, $x_F$, of the dimuon is
\begin{eqnarray}
x_F = \frac{P_L}{P_{\text{max}}} = \frac{2P_L}{\sqrt{s}\left(1-M^2/s\right)},
\label{def_xf}
\end{eqnarray}
where $P_L$ is the longitudinal momentum of the dimuon and $P_\textrm{max}$ is the maximum momentum in the center-of-mass frame of the colliding hadrons. 
With the definition of $x_F$ adopted in Eq.~\ref{def_xf}, the full range of $\left|x_F\right| < 1$ is covered. In contrast, for the alternative definition of $x_F = 2P_L/\sqrt{s}$, frequently used in the
literature, the coverage of $x_F$ is limited to $\left|x_F\right| <\left(1-M^2/s\right)$.
These two definitions of $x_F$ converge at the high-energy limit when $M^2/s \to 0$. It is worth noting that Eq.~\ref{def_mass} becomes the familiar expression $M^2=x_1 x_2 s$, often adopted in the literature, when $P_T/\sqrt{s} \to 0$. 
For the SeaQuest experiment, the modest beam momentum of 120 GeV $\left(\sqrt{s} = 15.1\text{~GeV}\right)$ warrants the use of the definitions of kinematic variables shown in Eqs.~\ref{def_mass} and \ref{def_xf}.

The leading order Drell-Yan cross section, expressed in Eq.~\ref{eq:DYCross}, contains two terms since the annihilation could proceed with the antiquark from either parent hadron. 
For most fixed-target experiments, including SeaQuest, the spectrometers have large acceptance for the positive $x_F$  region $\left(x_F > 0\right)$. Thus, to a good approximation, the Drell-Yan cross section is dominated by the first term, corresponding to the annihilation of a beam quark with a target antiquark. To demonstrate the relation between the cross section ratio and the flavor asymmetry $\bar{d}\left(x\right) / \bar{u}\left(x\right)$, an approximate formula can be derived as follows. Taking into account the dominance of $u\left(x\right)$ over $d\left(x\right)$ in the beam proton and the charge squared weighting factor $e_i^2$ in Eq.~\ref{eq:DYCross}, one obtains for $x_1 \gg x_2$,
\begin{align}
\frac{\sigma^{pd}}{2\sigma^{pp}} = \frac{1}{2}\left[1 + \frac{\sigma^{pn}}{\sigma^{pp}}\right]
\approx \frac{1}{2}\left[1 + \frac{\bar{u}_n\left(x_2\right)}{\bar{u}_p\left(x_2\right)}\right],  
\end{align}
with the assumption that $\sigma^{pd} \approx \sigma^{pp} + \sigma^{pn}$, which neglects small nuclear effects of the deuteron~\cite{Kamano,Ehlers}. Charge symmetry for the parton distributions~\cite{Londergan} between the proton and the neutron:
\begin{eqnarray}
\bar u\left(x\right) & \equiv & \bar u_p\left(x\right) = \bar d_n\left(x\right)\textrm{, and }\\
\bar d\left(x\right) & \equiv & \bar d_p\left(x\right) = \bar u_n\left(x\right),
\label{eq:chargeSymmetry}
\end{eqnarray}
leads to the approximate expression
\begin{eqnarray}
\frac{\sigma^{pd}}{2\sigma^{pp}} \approx
\frac{1}{2} \left[1+\frac{\bar d\left(x_2\right)}{\bar u\left(x_2\right)}\right].
\label{eq:crRatio}
\end{eqnarray}
While Eq.~\ref{eq:crRatio} illustrates the power of the $\sigma^{pd}/2\sigma^{pp}$ Drell-Yan cross section ratio to reveal the flavor asymmetry between $\bar d$ and $\bar u$, the actual extraction of the $\bar d\left(x\right) / \bar u\left(x\right)$ ratios from the measured $\sigma^{pd}/ 2 \sigma^{pp}$ Drell-Yan cross section ratios is made without these simplifying approximations and is carried out in Next-to-Leading Order (NLO) in the strong coupling constant, $\alpha_s$, as discussed in Sec.~\ref{sec:extraction}.

%this two column float gets stuck and placed later.  I moved it up so that is could appear in the page with Sec III the SeaQuest Experiment
\begin{figure*}[tb]
    \centering
    \includegraphics[angle=10,width=0.9\textwidth] {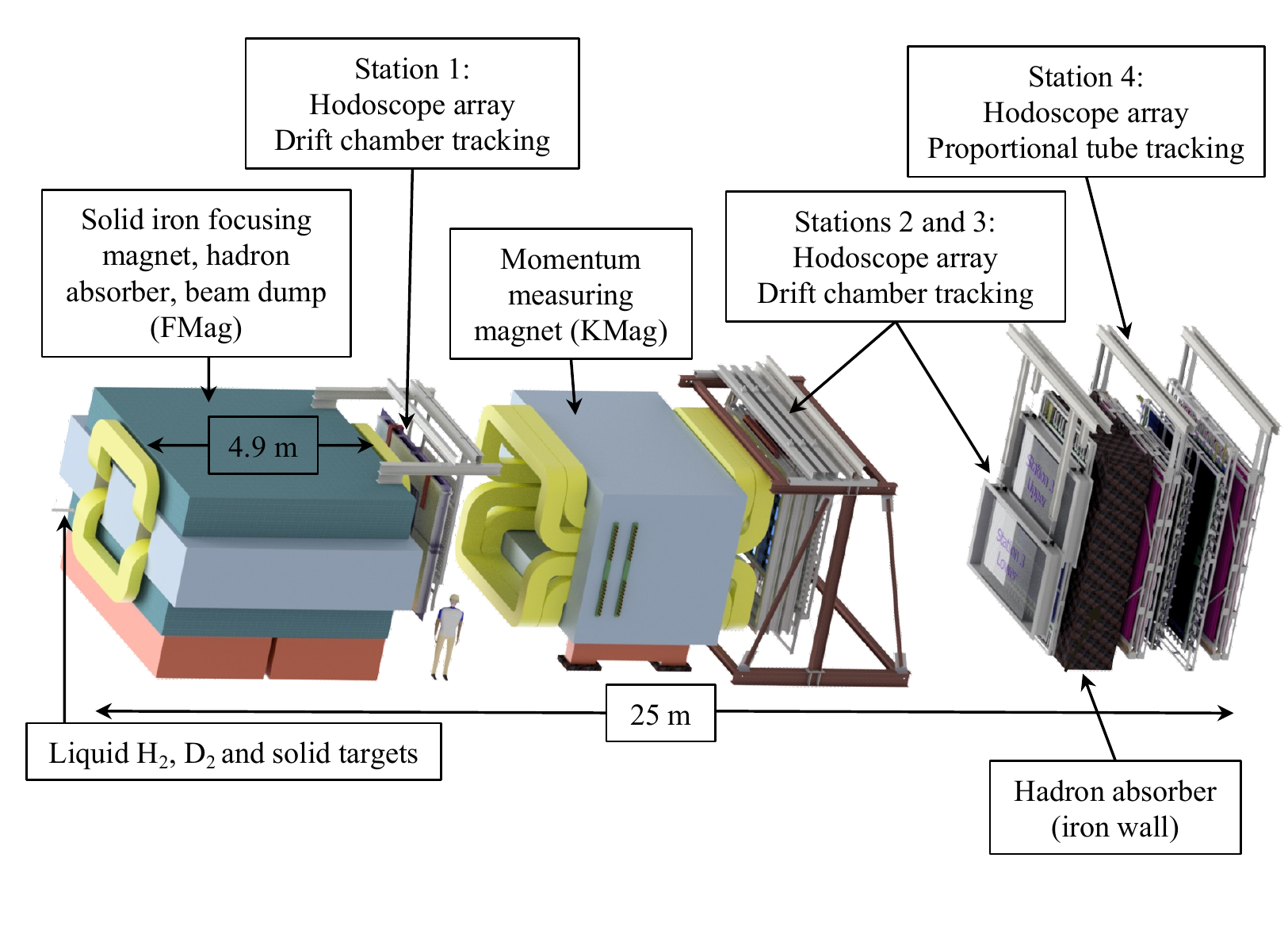} \vspace*{-1.0cm}
    \caption{Schematic of the SeaQuest spectrometer. \label{fig:SeaQuestApparatus}}
\end{figure*}

\section{The SeaQuest Experiment\label{sec:apparatus}}

The SeaQuest experiment uses a proton beam extracted from the Fermilab Main Injector at 120 GeV once every minute for a 4-second period (spill).  The beam inherits the 53.1 MHz RF structure from the accelerator which groups the protons into about 2 ns ``buckets" every 18.8 ns.  The average intensity of the beam is approximately $2 \times 10^{12}$ protons per second, a rate roughly 10 times that of the predecessor E866 experiment. The protons in the Main Injector are peeled off using an electromagnetic septum over the course of the 4-second spill.  While the {\em average} intensity was roughly $10^4$ protons per bucket, the measured numbers varied up to a few times of $10^5$ protons.

Buckets with a large number of protons produce many hits in the detectors, readily satisfy the trigger requirements, and cause significant data acquisition (DAQ) dead time. Moreover, these events are typically too noisy to be effectively reconstructed.  In order to overcome this obstacle, a Beam Intensity Monitor (BIM), based on a gas Cherenkov counter, was installed upstream of the target. The BIM is capable of measuring the proton intensity in each RF bucket and vetoing buckets whenever the number of protons exceeds a programmable threshold.  
%The threshold varied as a function of time due to the gradual deterioration and occasional replacement of the reflecting mirror in the BIM, but was generally set around $\left(6.5\text{--}9.5\right) \times 10^{4}$ protons. 
The BIM threshold was generally set around $\left(6.5\text{--}9.5\right) \times 10^{4}$ protons, but varied over time due to the gradual deterioration and occasional replacement of the reflecting mirror in the BIM.

\begin{figure}[tb]
    \centering
    \includegraphics[width=0.7\columnwidth]{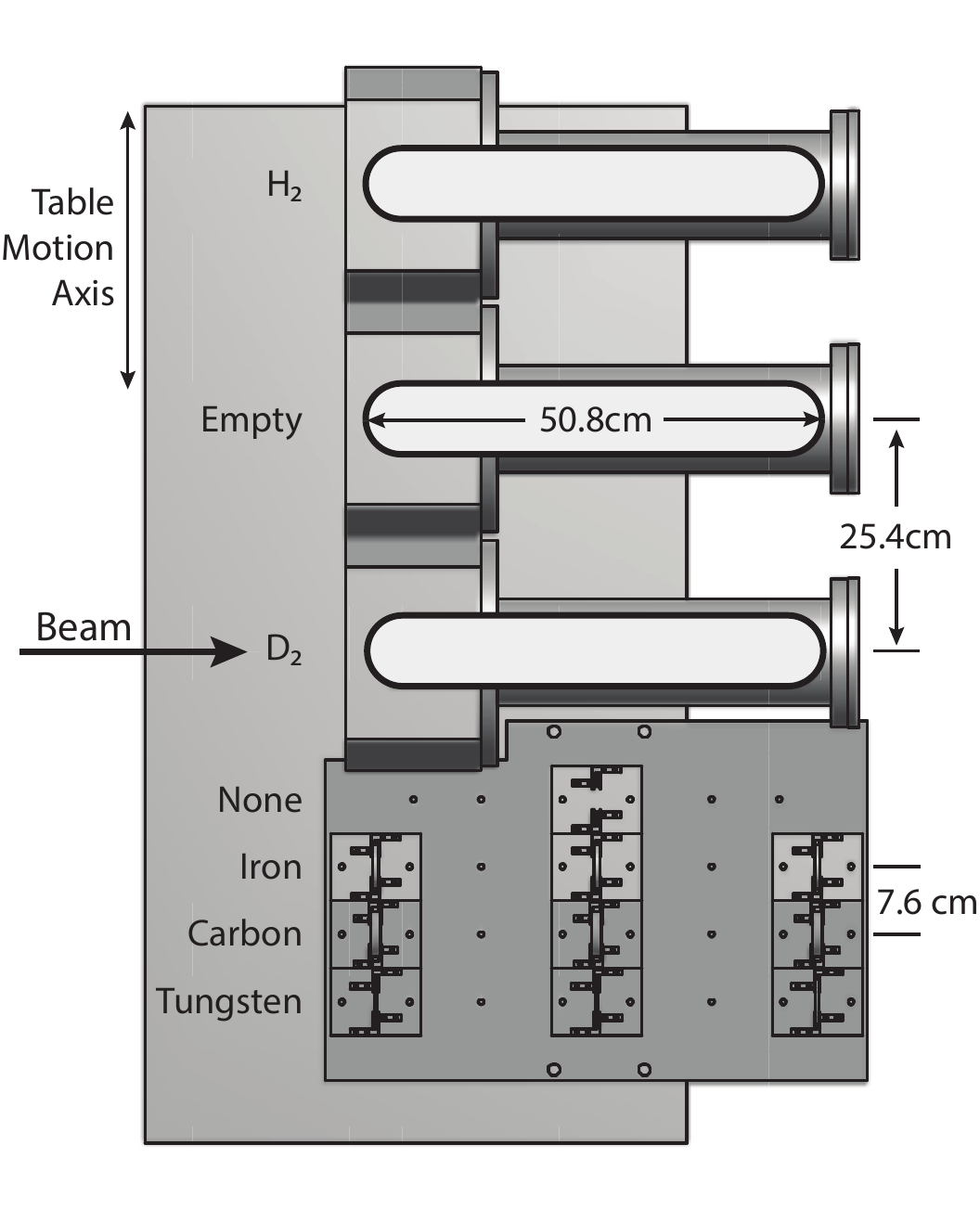}
    \caption{Schematic top view of the target table for SeaQuest. Between spills the table can shift horizontally to bring a new target into place. The frequent shifting of targets reduces systematic uncertainties caused by variations in the experimental conditions over time. \label{fig:target}}
\end{figure}

The schematic layout of the SeaQuest spectrometer, discussed in detail in Ref.~\cite{E906trig2}, is displayed in Fig.~\ref{fig:SeaQuestApparatus}. 
The proton beam is incident on one of the seven targets, two liquid targets (hydrogen and deuterium), three solid targets (carbon, iron, and tungsten) and two calibration targets (``empty flask" and ``no target").  
The two liquid targets are identical cryogenic cylindrical stainless steel flasks with hemispherical end caps.  
The flasks are 50.8 cm long with a 7.62 cm diameter containing 2.2 liters of liquid (Fig.~\ref{fig:target}).  
Each target flask has entrance and exit windows composed of a 51 $\mu$m-thick (0.002 in.) stainless steel end-cap of the flask and a 140 $\mu$m-thick (0.0055 in.) titanium window of the vacuum vessel that contains it.
The ``empty flask" calibration target is an evacuated flask identical to the flasks of the liquid targets, and the ``no target" is simply air.  
The targets are placed on a movable table which slides horizontally to switch targets in the 56-second period between two 4-second spills.  
Due to the different densities between the hydrogen and deuterium targets, the number of spills taken on the hydrogen target is roughly twice as large as on deuterium to acquire similar statistical precision for dimuon events for the two targets (see Tab.~\ref{tab:Targets}).

%\begin{table*}[tb]
\begin{table}
\caption{Density and thickness of the target material and number of spills per target-rotation cycle for the SeaQuest targets used in the present analysis, where ``D'' denotes ``$^2$H''.}
%\caption{Physical parameters and number of spills per target-rotation cycle for SeaQuest targets \bc {used in the present analysis}.}
\begin{tabular}{lcclc}
\hline\hline
 & Density  & Thickness & Interaction & \\
Material    &  (g/cm$^3$) &  (cm) & Lengths & Spills/Cycle \\
\hline
H$_2$ & 0.071 & 50.8 & 0.069 & 10\\
Empty Flask & - & - & 0.0016 & 2\\
D$_2$ & 0.162 & 50.8 & 0.115 & 5            \\
           \hline\hline
\end{tabular}
\label{tab:Targets}
\end{table}

Approximately 130 cm downstream of the target, the beam encounters a large, solid-iron dipole magnet called FMag (Fig.~\ref{fig:SeaQuestApparatus}).   
The right-handed coordinate system has the $z$-axis pointing along the beam direction and the $y$-axis is oriented upward in the vertical direction. 
FMag has a size of 5.0 $\times$ 3.0 $\times$ 4.9 meters.
It serves as a focusing magnet, a hadron absorber, and a beam dump. 
A vertical 2.07 Tesla magnetic field in FMag provides a 3.07 GeV transverse momentum kick to the muons passing through the FMag.

The second dipole magnet (KMag) is an open aperture magnet repurposed 
from the Fermilab E799/KTeV experiment.  It has an aperture of 
3 $\times$ 2.9 $\times$ 2 meters with a 0.39 Tesla magnetic
field in a direction
parallel to that of FMag, resulting in an additional 
0.39 GeV transverse momentum kick. The KMag further 
focuses the muon tracks and allows their momenta to be measured.  

The SeaQuest spectrometer consists of four detector stations.  Station 1 is positioned between FMag and KMag.  It consists of 
two scintillator hodoscope planes and drift chambers.  
The hodoscope planes are 
oriented in the $x$ and $y$ directions. The  
hodoscopes provide a spatially coarse but temporally fast measurement for the trigger system.  
Immediately following the hodoscope panels is a drift chamber 
consisting of three pairs of wire planes oriented vertically 
and $\pm 14^\circ$ with respect to the vertical.  
Stations 2 and 3, located downstream of the KMag, are approximately 
5 meters apart.  
Station 4, situated roughly 2 meters behind station 3 with 
a 1-m thick hadron-absorbing iron wall in between, is primarily used for muon identification.

The SeaQuest FPGA-based trigger system is described in detail in 
Refs.~\cite{E906trig1,E906trig2}. Discriminated signals from the
hodoscope counters form the inputs for four CAEN V1495 FPGA VME modules,
one for each hodoscope ``quadrant" (upper and lower sections
of $x$-measuring hodoscopes, and upper and lower sections of 
$y$-measuring hodoscopes). The pattern of hodoscope hits in a given quadrant is compared with a lookup table of ``Trigger Roads" generated from Monte Carlo for valid muon tracks originating from the target region. Outputs of each of the four muon track finders, binned according to track charge and momentum,
are then checked against a lookup table of valid ``Dimuon-roads".
Triggers, defined as combinations of possible tracks with specific 
quadrant pairs, are listed in Tab.~\ref{tab:triggers}.  
Trigger 1 is the primary dimuon trigger, recording pairs of unlike-sign tracks 
likely to be produced at the target region.  
%Trigger 2 is designed for dimuons with larger transverse momenta. 
%Trigger 3, the like-sign trigger, directly measures the coincidence of two tracks with same charge signs. 
Trigger 4 records single tracks and was used to evaluate the accidental coincidence background formed from the 
combination of two uncorrelated tracks. 
In addition to these triggers, there is a trigger with random timing intended for sampling the structure of the beam and the response of the detectors.

\begin{table}[tb]
\caption{Triggers used in the present analysis. The T or B denotes the top or bottom section traversed by the track. \label{tab:triggers}}
\begin{tabular}{c@{\hspace{6\tabcolsep}}c@{\hspace{6\tabcolsep}}c@{\hspace{6\tabcolsep}}l}
\hline
\hline
Trigger     & Side  & Charge & Description                   \\
\hline
1 & TB/BT & $+-/-+$                     & Unlike-sign dimuon     \\
4 & T/B   & $+/-$                      & Single muon      \\
\hline
\hline
\end{tabular}
\end{table}

\section{Data Analysis\label{sec:dataAnalysis}}

This analysis includes data recorded between June 2014 and July 2015, containing roughly half of the total data collected in SeaQuest. The events used in this analysis were collected with the dimuon trigger (trigger 1).  The analysis reconstructed the trajectories and momenta of charged muons, and selected candidate events consisting of a pair of $\mu^+\mu^-$ originating from the target region. 
The stability of reconstructed quantities over the data-taking condition has been examined, and spills that showed large variations were studied and excluded if issues were found.

\subsection{Track Reconstruction \label{sec:tracking}}

Track reconstruction begins with the removal of spurious wire chamber hits, consisting of several different categories: out-of-time hits, after-pulse hits, and cluster hits.  The out-of-time hits have a drift time outside of an allowable range.  The after-pulse hits are caused by occasional ringing in the readout electronics.  Cluster hits, originating from delta rays or cosmic rays travelling parallel to the detector plane, result in three or more hits in adjacent wires.  In addition, chamber hits not matched by hits in the geometrically  correlated scintillator hodoscope are removed.

After removing these hits, the remaining chamber hits are organized into straight-line segments (called tracklets) for each of the first three stations. The tracklets are matched to find valid tracks consistent with originating from the target region. A Kalman filtering procedure is then applied to find the best estimate for the momentum and trajectories of the track, taking into account the multiple scattering and energy loss of the muons passing through the solid iron FMag. Finally, pairs of tracks in the top and bottom sections are combined and another Kalman filter is performed to construct the momentum and vertex position of the muon pairs. 

Additional analysis cuts are applied to select the final dimuon  candidates.  These analysis cuts can be separated into four categories: track cuts, dimuons cuts, physics cuts, and intensity cuts.  Track cuts are applied on single track variables, such as the momentum or position of each track.  Dimuon cuts are made on the dimuon quantities, such as the vertex position or dimuon momentum.  Physics cuts are applied on reconstructed physics variables such as dimuon mass $\left(M\right)$ or $x_F$.  Finally, intensity cuts are made on the intensity of the event, either in the measured proton intensity from the BIM, or the wire chamber occupancy.

\begin{figure}
    \centering
    \includegraphics[width=0.9\columnwidth]{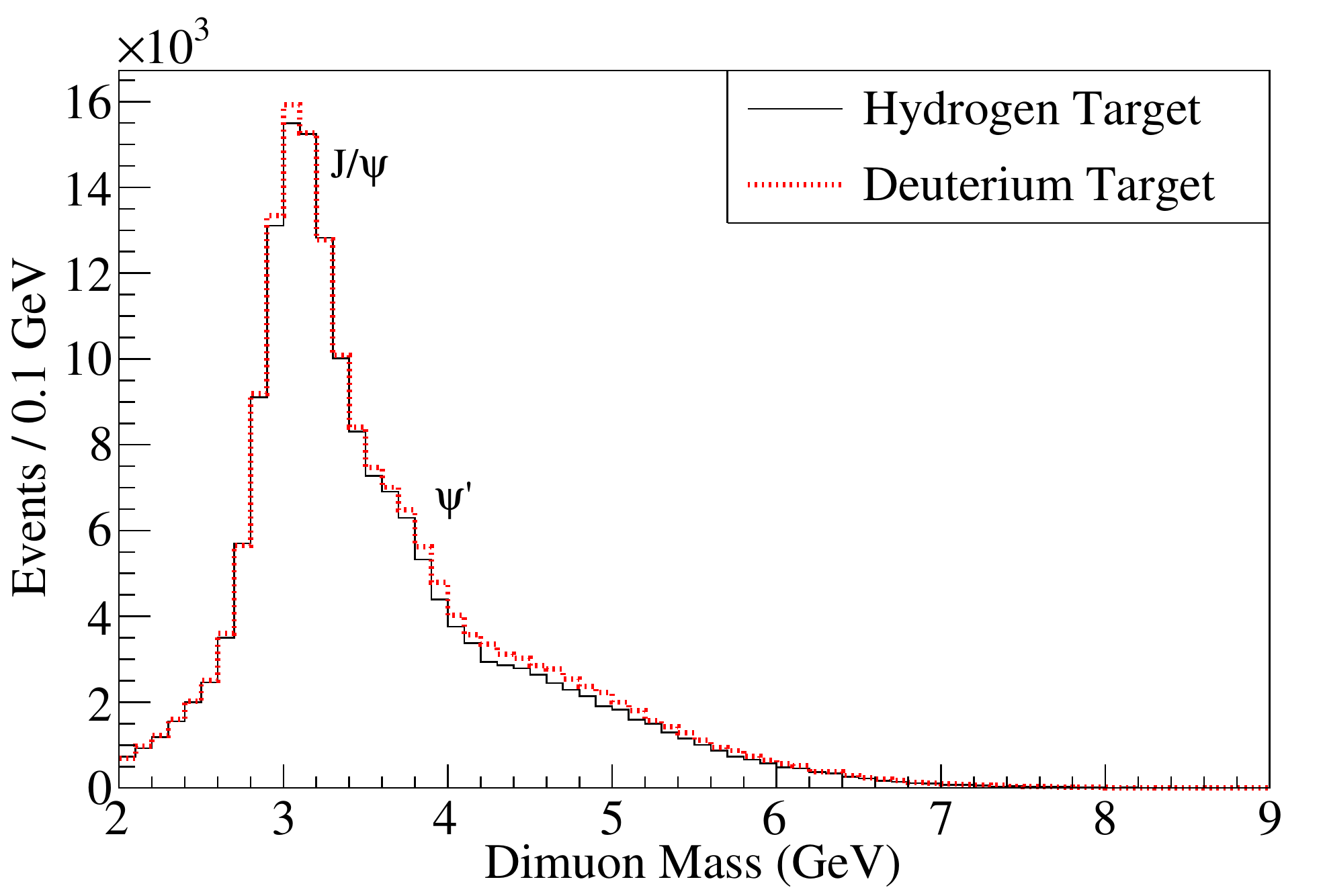}
    \caption{Mass spectrum of dimuon candidate events from interaction
of proton beam with the liquid hydrogen and deuterium targets.  
The $J/\psi$ peak and $\psi'$ shoulder are visible with 
the Drell-Yan continuum extending to the high mass region.
}
    \label{fig:mass_spectrum_LD2}
\end{figure}

The dimuon mass distribution, shown in Fig.~\ref{fig:mass_spectrum_LD2} for both hydrogen and deuterium targets has several prominent features. The $J/\psi$ peak is clearly observed and the $\psi^\prime$ shoulder is visible. The dimuon continuum extends to $M\approx 9$ GeV. A small number of dimuon events, originating from the interaction of the beam with the flask windows or other materials upstream or downstream of the liquid target, is measured using the empty-flask target. 
In addition to true dimuons, the presence of background from the coincidence two muons produced in separate interactions is expected. 
This background was subtracted by using the events collected with the single-muon trigger (trigger 4), as discussed in Sec.~\ref{sec:crossRatio}.

\begin{figure}
    \centering
    \includegraphics[width=\columnwidth]{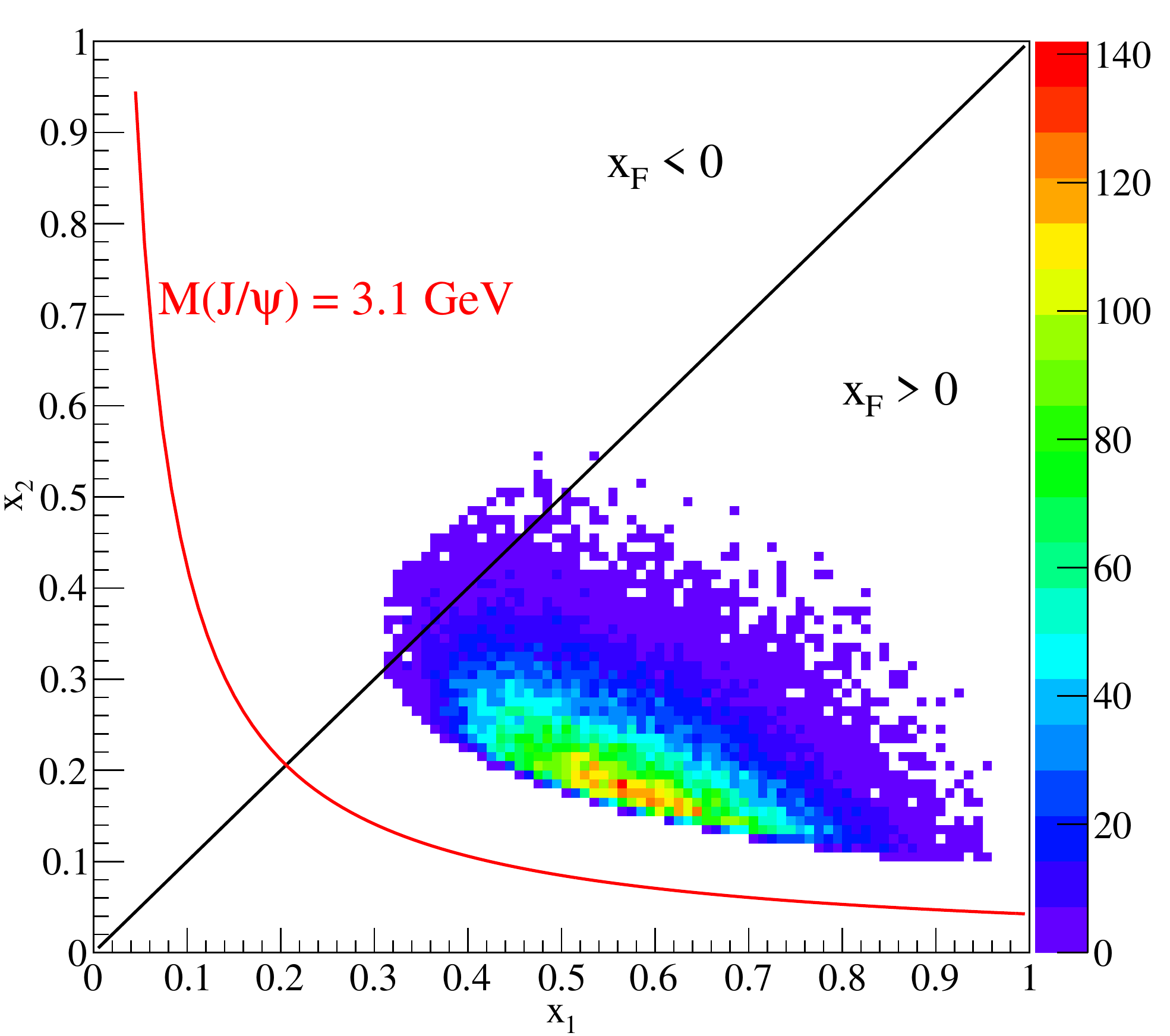}
    \caption{The $x_2$ versus $x_1$ event distributions of dimuon candidate events for $p+d$ after the $M > 4.5$ GeV cut. The black line corresponds to $x_F = 0$, and the red curve indicates the locations for $J/\psi$ kinematics (for $P_T=0$).
    \label{fig:acceptance_x12} }
\end{figure}

\begin{figure}
    \centering
    \includegraphics[width=0.49\columnwidth]{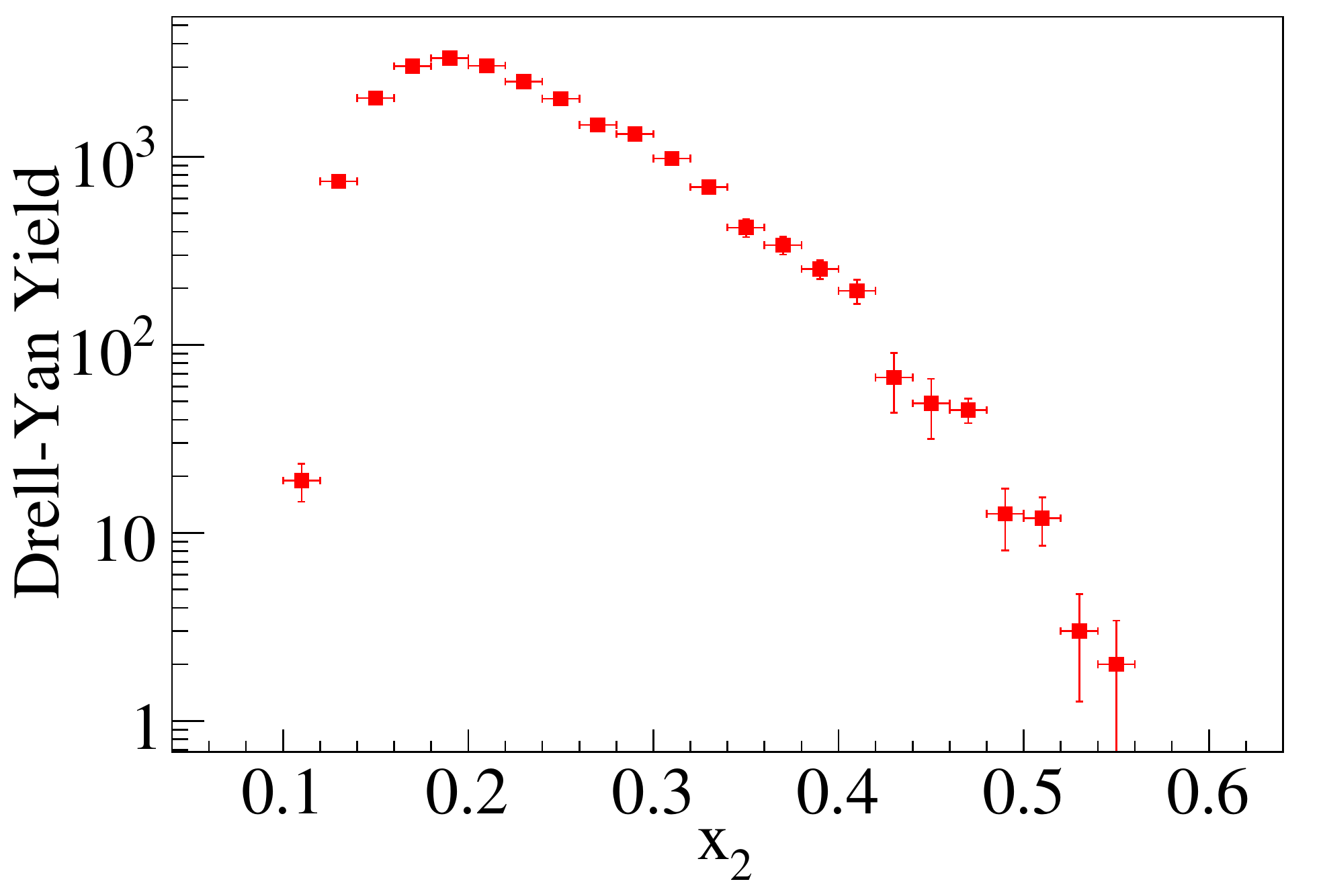}\hfill
    \includegraphics[width=0.49\columnwidth]{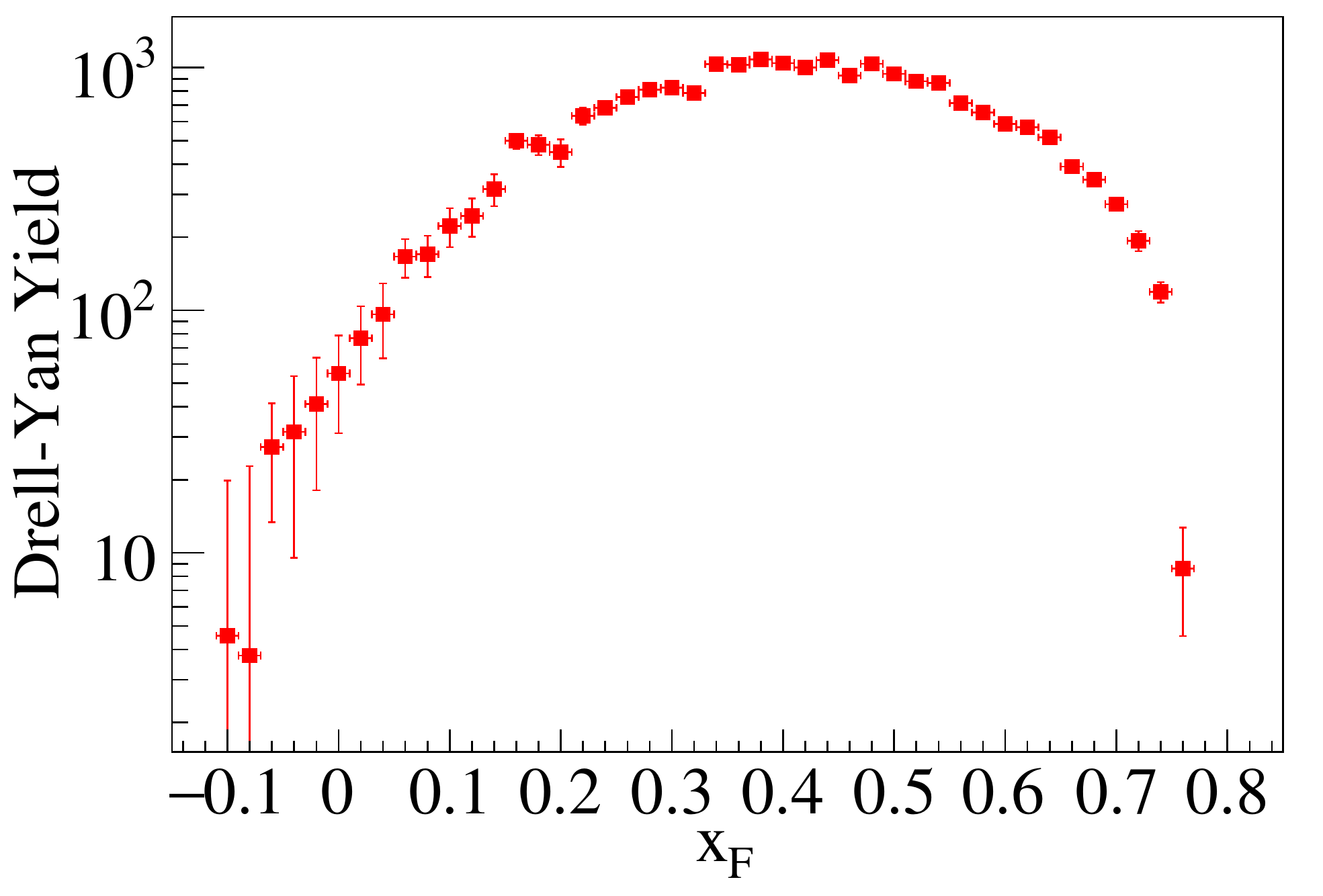} \\
    \includegraphics[width=0.49\columnwidth]{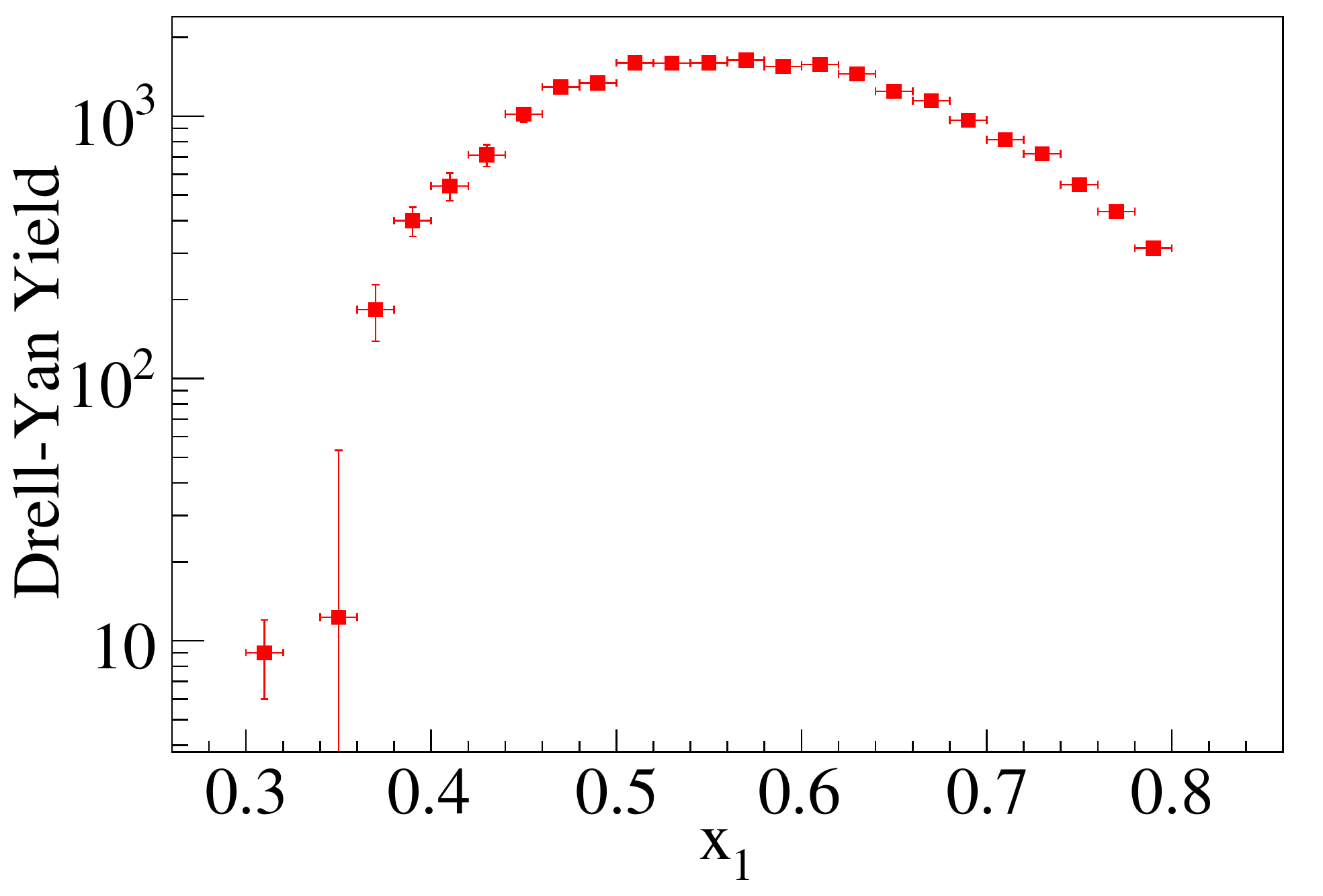} \hfill
    \includegraphics[width=0.49\columnwidth]{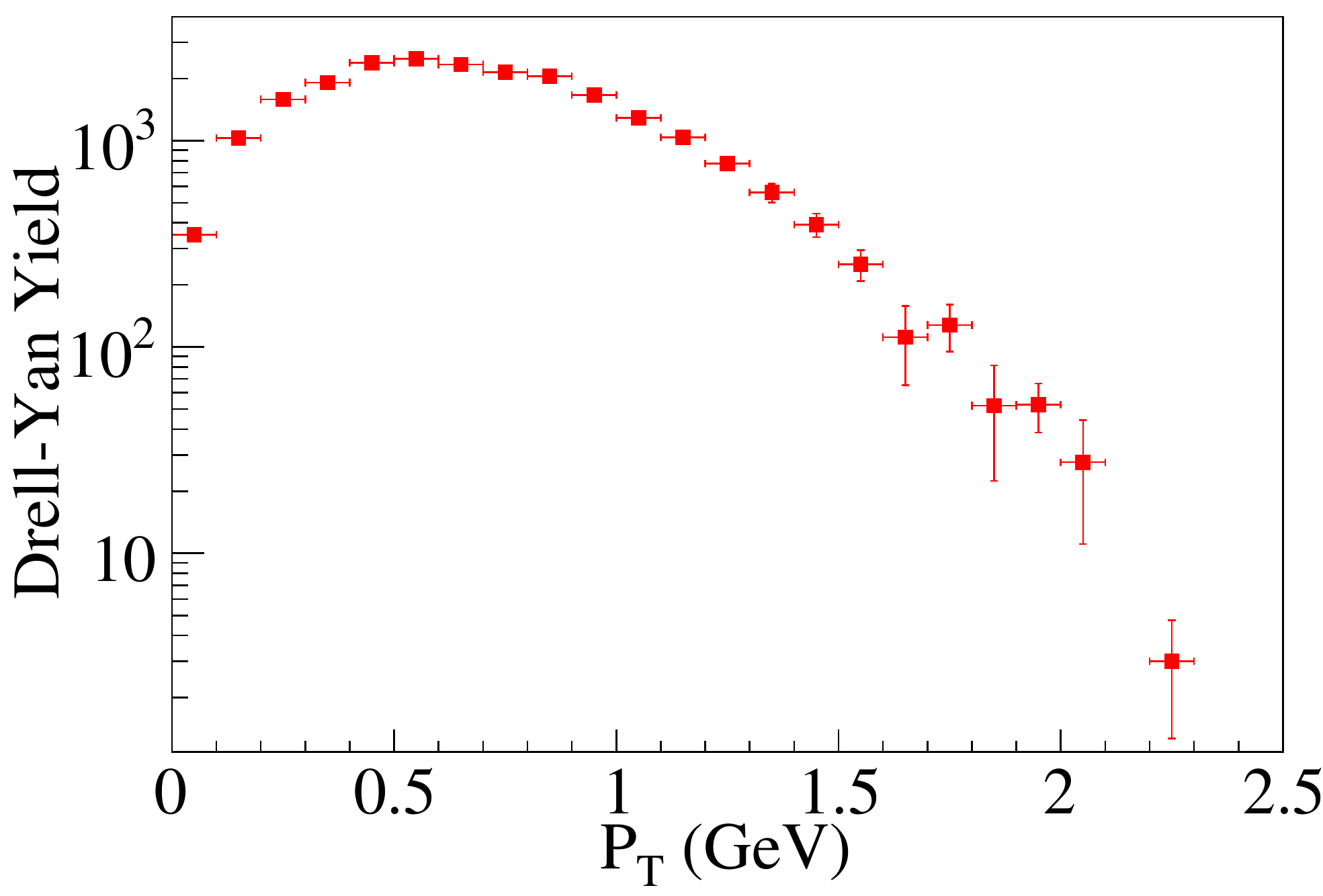} 
    \caption{The dimuon event spectra from $p+d$ interactions projected onto $x_2$, $x_1$, $x_F$, and $P_T$ variables. \label{fig:dimuon_dist_proj}}
\end{figure}

To examine the distributions of the Drell-Yan events versus various kinematic variables, a mass cut at $M > 4.5$ GeV is applied to remove the $J/\psi$ and $\psi^\prime$ events. 
Figure~\ref{fig:acceptance_x12} shows the $x_2$ versus $x_1$
distributions for the dimuon events after the mass cut. The
black line in Fig.~\ref{fig:acceptance_x12} represents $x_F = 0$, showing that the Drell-Yan candidate events are predominantly in the $x_F > 0$ region. From Eqs.~\ref{eq:DYCross} and \ref{eq:crRatio}, one expects the SeaQuest Drell-Yan data to be sensitive to the antiquarks in the target nucleons. Figure~\ref{fig:dimuon_dist_proj} shows the distributions of the $p+d$ dimuon events versus the kinematic variables $x_1$, $x_2$, $x_F$ and $P_T$ after the $M > 4.5$ GeV cut.

\subsection{Monte Carlo Simulation \label{sec:MonteCarlo}}
A Monte Carlo code was developed to compare the detector performance and experimental results with expectations.
Drell-Yan events were generated with a distribution in $M$ and $x_F$ based on an NLO cross section calculation using the active flavors $u$, $d$, $s$, and $c$.  
The $P_T$ distribution has been tuned so that kinematic distributions matched those observed in the measured data.  
The CT14 PDFs~\cite{Dulat:2015mca} were used in these calculations.  
As a check on systematic effects, the calculations were also repeated with PDFs from a statistical model~\cite{Basso}.

The $J/\psi$ and $\psi'$ Monte Carlos were based on a parameterized cross section extracted from experimental data~\cite{PhysRevD.52.1307};
\begin{equation}
BR\ \sigma^{J/\psi}(x_F) = A \exp(-B M / \sqrt{s}) \frac{\exp(-{x_F}^2 / W^2)}{\sqrt{2\pi} W},
\end{equation}
where $M = 3.097$ GeV, $A = 1464$ nb, $B = 16.66$, $W = 0.2398$ and $BR = 0.0594$.
For $\psi'$, the cross section was scaled with the yield ratio of $\psi'$ to $J/\psi$ in dimuon channel, $\alpha = 0.019$;
\begin{equation}
BR\ \sigma^{\psi'}(x_F) = \alpha\ BR\ \sigma^{J/\psi}(x_F).
\end{equation}
Event yields have been re-weighted so that kinematic distributions matched with real data.

The entire apparatus of the SeaQuest spectrometer was modeled and the response to dimuons was computed, 
based on Geant4-9.6.2~\cite{GEANT4:2002zbu,1610988,ALLISON2016186}.
The interaction of muons with the target and spectrometer materials was simulated using a physics list of {\tt FTFP\_BERT\_EMX}.
The energy loss and multiple scattering at FMag dominate the resolution of muon momenta.
Every energy deposit on detector plane was digitized to a hit on the corresponding element.
Drift chambers hits went through a ``realization'' step to add detector resolution through Gaussian smearing to the drift distance and inefficiency by dropping a small number of randomly selected hits. 
Both the width of the smearing and the efficiency were obtained from the data.

Noise hits were added to each Monte Carlo event by embedding hits from events that were recorded using the random trigger.
This step simulates background hits that arise from extra $p+p$ or $p+d$ reactions per RF bucket. 
It facilitates the determination of the tracking reconstruction efficiency as a function of chamber occupancy. 
It also allows us to evaluate the accuracy of the track reconstruction algorithm in obtaining true dimuon kinematics for high chamber-occupancy events.

The Monte Carlo events for the Drell-Yan, $J/\psi$, and $\psi^\prime$ dimuons following the realization and embedding procedures are analyzed with the algorithms and cuts used for the real data. 
Analyses using the Monte Carlo events are essential for the second method of extracting the cross section ratios  discussed in Sec.\ref{sec:MF}.

\section{Measurement of the $\sigma^{pd}/2\sigma^{pp}$ Drell-Yan cross section ratios\label{sec:crossRatio}}

Ideally, the Drell-Yan event rate for a given target should be linearly proportional to beam intensity.
However, there are several experimental effects that can introduce a nonlinear dependence of the reconstructed dimuon event rate on beam intensity.
For example, the coincidence between two muons produced
in two different interactions will yield a contribution that scales quadratically with the beam intensity. In addition, inefficiencies from the dead time of the DAQ
system, as well as from the event reconstruction, can  produce
non-linear dependencies of the dimuon event rates on beam intensity.
There is also a 
significant variation of beam intensity during a beam spill, causing a
time-dependence for these effects. These rate-dependent effects are expected to be larger for the deuterium target than for the hydrogen target, because the density of the deuterium target is higher than that of the hydrogen target, and will affect the measured ratios. Two different analysis methods have been developed by SeaQuest to take these effects into account.   
The Intensity Extrapolation (IE) method, which was described in some
detail in a recent paper~\cite{Dove21}, examines the beam intensity dependence of the dimuon yield to extract the ratio of the Drell-Yan signal events. In this paper, we also present an independent approach, the Mass-Fit (MF) method, to determine various dimuon components from a fit to the mass spectra. These components can then be removed from the event yields for the hydrogen and deuterium targets, leaving the number of Drell-Yan events. In this section we discuss these methods and compare the results of these two approaches. 

%\begin{figure}
%\includegraphics[width=0.48\textwidth]{figs/ieall.pdf}
%    \caption{Ratio of the deuterium to hydrogen Drell-Yan events as a function of intensity.  The solid curve shows the fit using a quadratic function according to Eq.~\ref{eq:ComPol2}.}
%    \label{fig:intensityExtrap_integrated}
%\end{figure}

\subsection{Intensity Extrapolation Method}

\begin{figure}
    \includegraphics[width=0.48\textwidth]{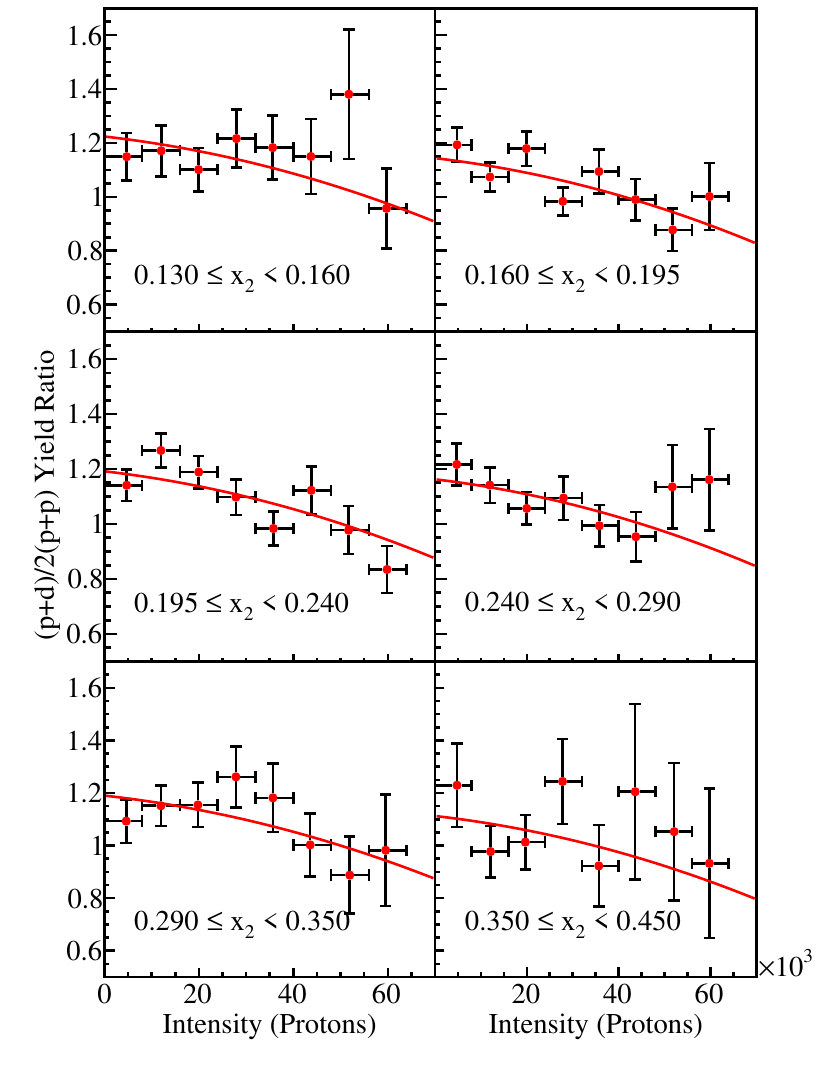}
    \caption{Ratio of $(p+d)/2(p+p)$ Drell-Yan events as a function of proton intensity for individual $x_2$ bins. The solid curve shows the fit using a quadratic function according to Eq.~\ref{eq:ComPol2}.  Comon systematic uncertainties are not shown.
    \label{fig:extrap} }
\end{figure}

In the IE method, the ratio of the $\left(p+d\right)$ to $2\left(p+p\right)$ dimuon event rates is evaluated as a function of the instantaneous beam intensity. By extrapolating these ratios to zero intensity, the $\sigma^{pd}/2\sigma^{pp}$ Drell-Yan cross section ratio, called $R$, is determined. In this method, several dimuon kinematic cuts, including $M > 4.5$ GeV, $x_1 < 0.8$, and $x_F > -0.1$, are applied to the $p+p$ and $p+d$ dimuon candidate events to remove the charmonium events and to exclude regions close to the boundaries of the spectrometer acceptance. Data collected with the empty flask target, normalized by the integrated beam intensity, determine the number of events in the $\left(p+d\right)$ and $\left(p+p\right)$ that were produced in the flask.  
The $\left(p+d\right) /2 \left(p+p\right)$ ratios are formed from the empty-flask subtracted dimuon data, again normalized by the integrated beam intensity and the amount of $H_2$ and $D_2$ for each liquid target.
These ratios are binned as a function of the instantaneous beam intensity measured by the BIM for the specific RF bucket that triggered the DAQ system. The implementation of the BIM detector in the SeaQuest experiment is essential for providing the beam-intensity information for the triggering RF bucket. Figure~\ref{fig:extrap} shows the $\left(p+d\right) / 2\left(p+p\right)$ ratios of the dimuon events versus the proton intensity of the triggering RF bucket. In the absence of non-linear dependencies on the beam intensity, a constant ratio is expected. The negative slopes observed in Fig.~\ref{fig:extrap} reflect the combined effects of accidental coincidence, DAQ dead time, and reconstruction inefficiency described in Sec.~\ref{sec:dataAnalysis}. The final step in the IE method is to fit the intensity distributions of the $\left(p+d\right) / 2\left(p+p\right)$ ratios. The intercept of the fitted function then gives the value of $R$, the $\sigma^{pd}/2\sigma^{pp}$
Drell-Yan cross section ratio.

In order to obtain the $x_2$ dependence of $R$, the data are split into 6 separate $x_2$ bins and a second-order polynomial of the form
\begin{equation}
    R_i\left(I\right) = p_{0i}+p_1I+p_2I^2
    \label{eq:ComPol2}
\end{equation}
was used to fit the intensity dependencies, where $i$ denotes the $x_2$ bin number. The variables $p_1$ and $p_2$ are common to all bins, while the intercepts $p_{0i}$ give the values of $R$ for each $x_2$ bins. 
All the parameters were obtained by one simultaneous fit.
The result of the fit is shown in Fig.~\ref{fig:extrap}.  
The intensity dependence does not have an analytical form, but rather is a product of multiple effects including random backgrounds and detector efficiencies.  Thus we approximated it with the Taylor expansion.  We have also tried a third-order polynomial but observed an over-fit, namely the fit result was pulled by point-by-point fluctuations.
Other fitting functions were used to estimate the systematic uncertainties associated with the specific functional form chosen for the fitting.

\begin{figure} \centering
    \includegraphics[width=\columnwidth]{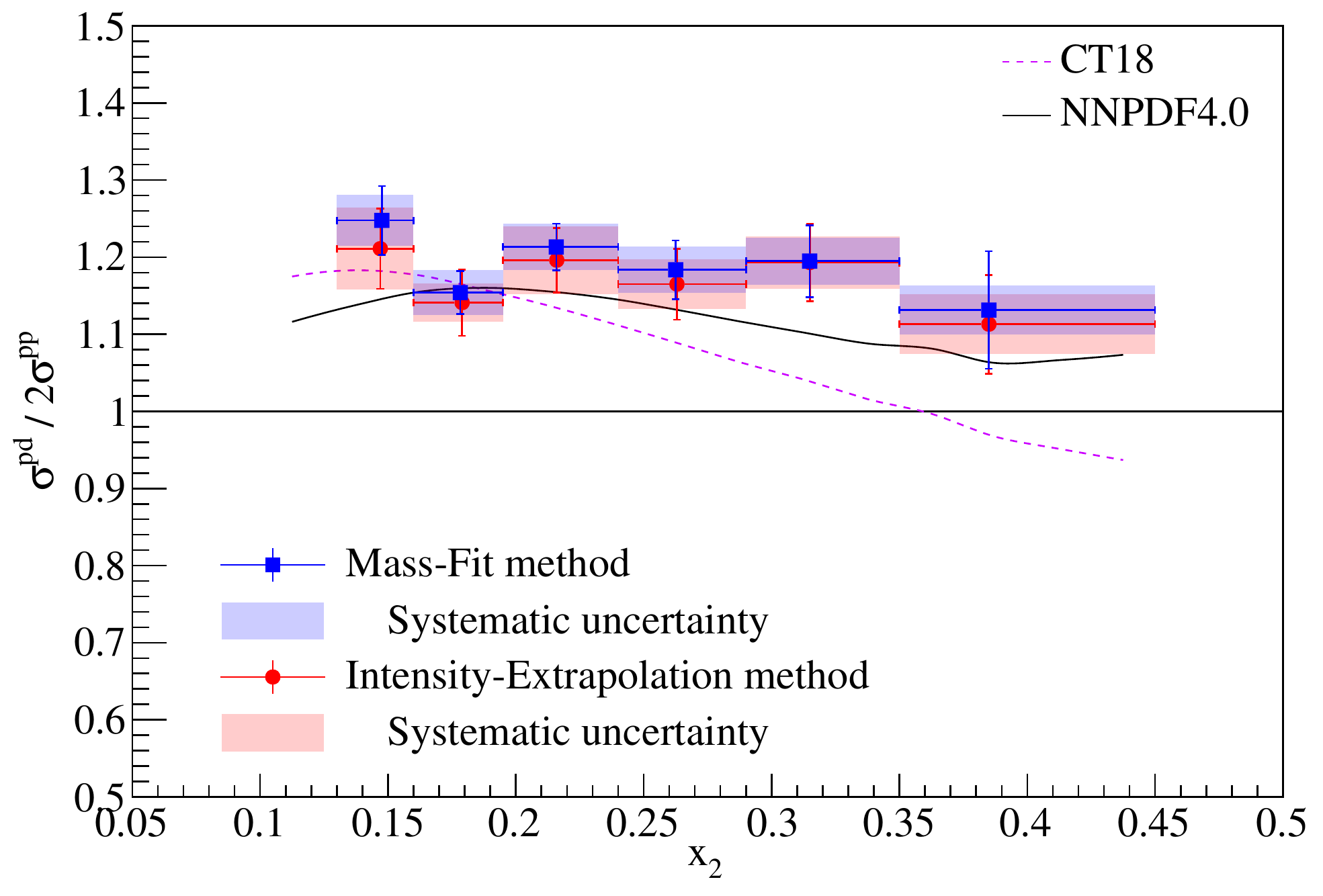}
    \includegraphics[width=\columnwidth]{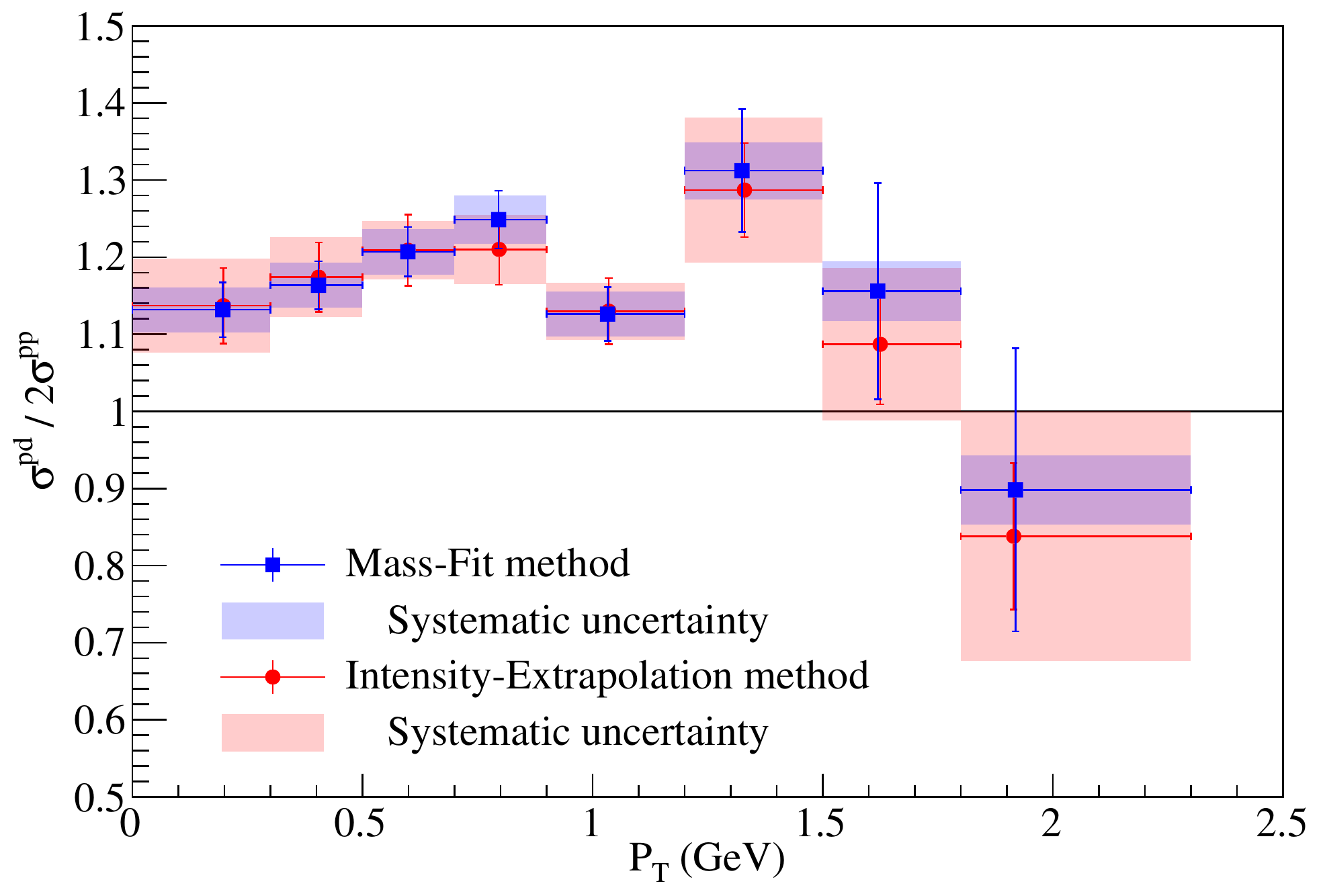}
    \caption{The cross section ratios versus $x_2$ (top) and $P_T$ (bottom) obtained with the two analysis methods.  
    The curves represent calculations using two different proton parton distributions, the CT18 (red, dotted) and the NNPDF4.0 (black, solid), weighted by the SeaQuest acceptance.
    \label{fig:CSR} }
\end{figure}

The IE method had been used to extract $R$ as a function of the kinematic variables $x_2$ and $P_T$ of the dimuons. The results were reported in Ref.~\cite{Dove21} and are displayed in Fig.~\ref{fig:CSR}.

\subsection{Mass-Fit Method \label{sec:MF}}

The Mass-Fit (MF) method was developed as an independent analysis to verify the results obtained with the IE method. The MF method can also be used to extract other quantities of interest, including the absolute Drell-Yan, $J/\psi$, and $\psi^\prime$ cross sections on the individual hydrogen or deuterium target.

\begin{figure} \centering
 \includegraphics[width=0.95\columnwidth]{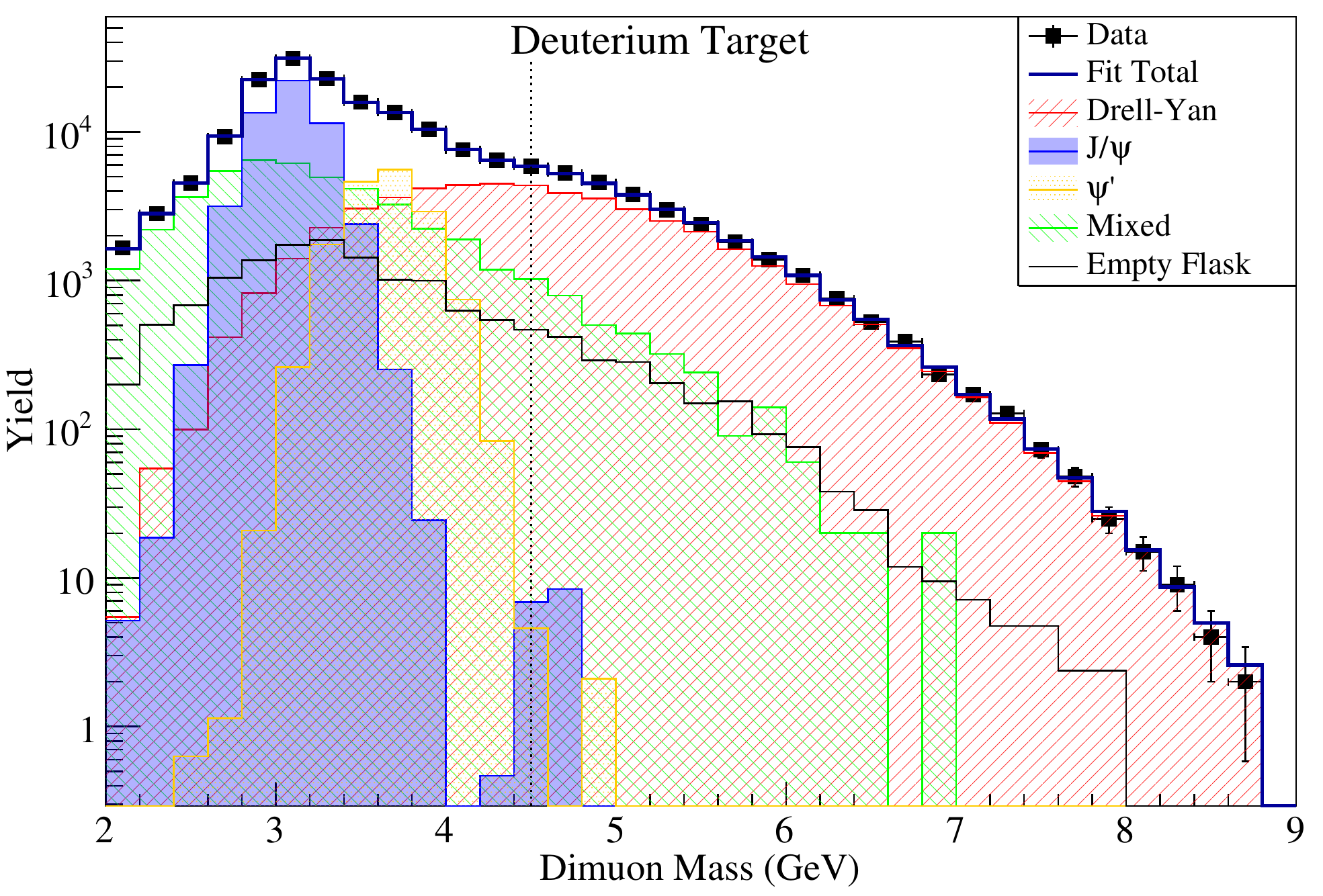}
 \includegraphics[width=0.95\columnwidth]{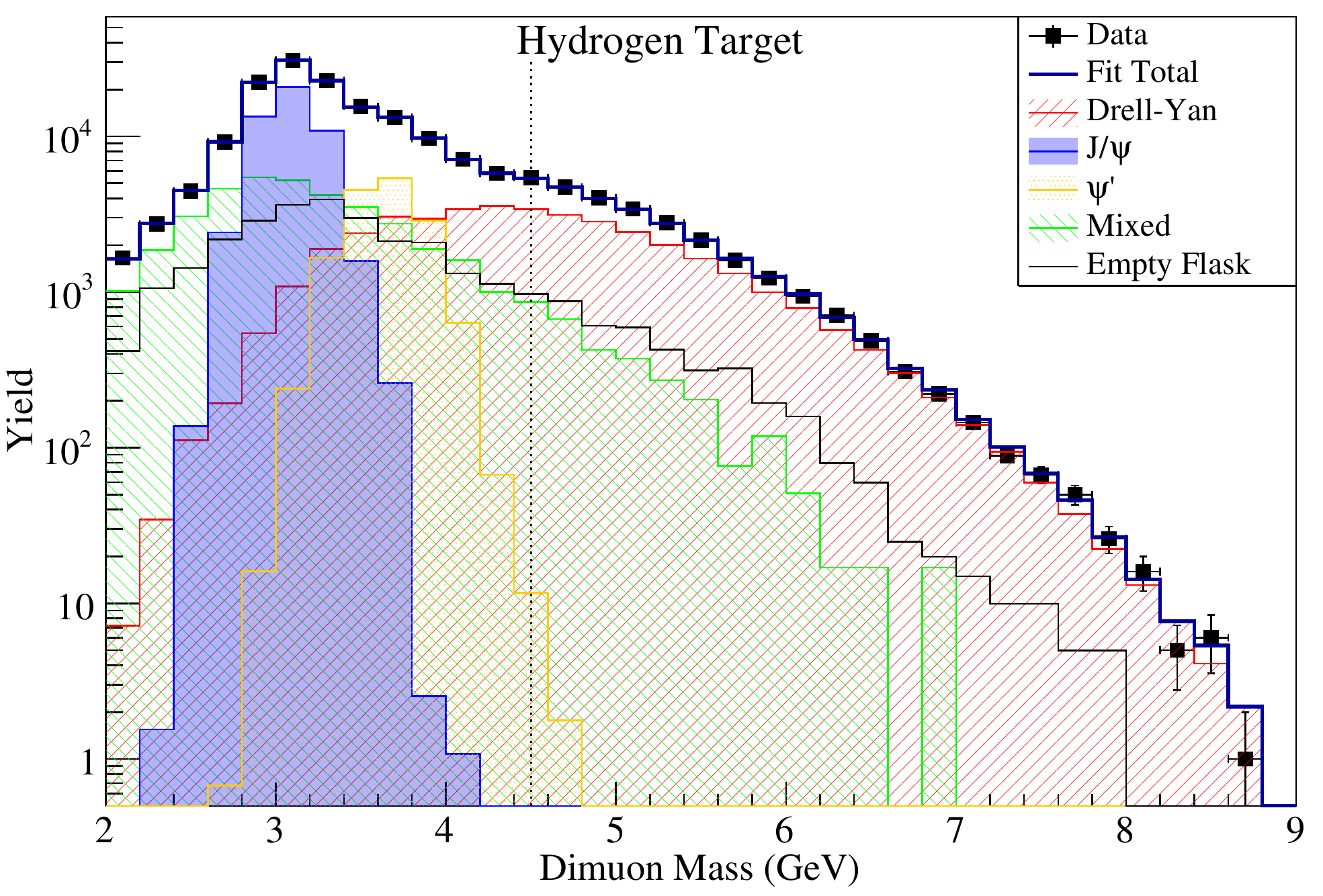}
    \caption{Mass fits of the deuterium (top) and the hydrogen (bottom) data after a majority of the analysis cuts.  The mass, $x_2$, and $x_1$ cuts are not applied in order to include the data at the low mass region for better determination of the normalization of the accidental background.  The vertical dotted line indicates $M = 4.5$ GeV. \label{fig:massFit_LD2} }
\end{figure}

The dimuon mass spectrum shown in Fig.~\ref{fig:mass_spectrum_LD2} consists of several components including Drell-Yan, $J/\psi$, $\psi'$, empty flask, and accidental coincidence events. To extract the Drell-Yan events, a fit to the dimuon mass spectrum was performed to determine the contributions from various components. As input parameters for the fit,  the mass distributions of the Drell-Yan, $J/\psi$, and $\psi'$ events were made with the Monte Carlo simulation described in Sec.~\ref{sec:MonteCarlo}.  The spectrum of accidental coincidence events was made by pairing at random $\mu^+$ and $\mu^-$ collected with the ``single-muon" trigger (trigger 4 of Tab.~\ref{tab:triggers}) under the condition that the beam intensities of $\mu^+$ and $\mu^-$ events were comparable.
%Then, the J/$\psi$, $\psi'$, and Drell-Yan Monte Carlo events, together with the empty-flask data and the accidental background, were analyzed in the same fashion as the data for the liquid targets.
The magnitudes of these components, 
with the exception of the empty-flask data for which the normalization is known, 
were varied in the fit to the mass spectrum. 
Figure \ref{fig:massFit_LD2} shows the fits to the $p+d$ and $p+p$ dimuon spectra, respectively. 
The distinct shapes for the mass spectra of the various components facilitate the decomposition of the measured dimuon events into various components.
The fit was redone with several input parameters varied, such as the PDF used in the simulations and the ranges of analysis cuts.
The effect on the cross section ratios was found small and included in the systematic uncertainty. 

%\begin{figure}
% \centering
% \includegraphics[width=0.5\textwidth]{h1_mass_fit_tgt1.pdf}
%    \caption{Same as Fig.~\ref{fig:massFit_LD2}, but for the hydrogen target.}
%    \label{fig:massFit_LH2}
%\end{figure}

\begin{figure} \centering
    \includegraphics[width=\columnwidth]{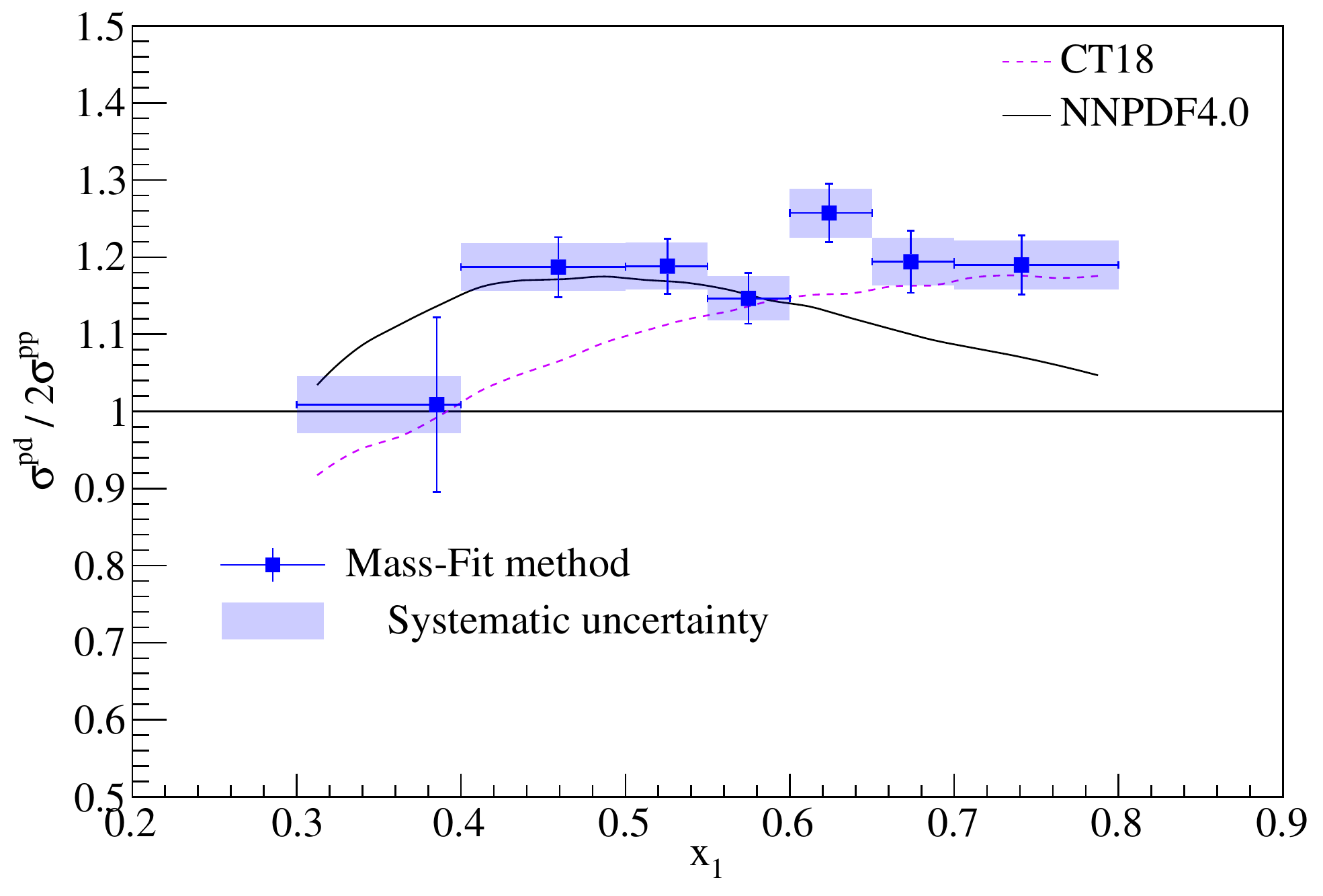} 
    \includegraphics[width=\columnwidth]{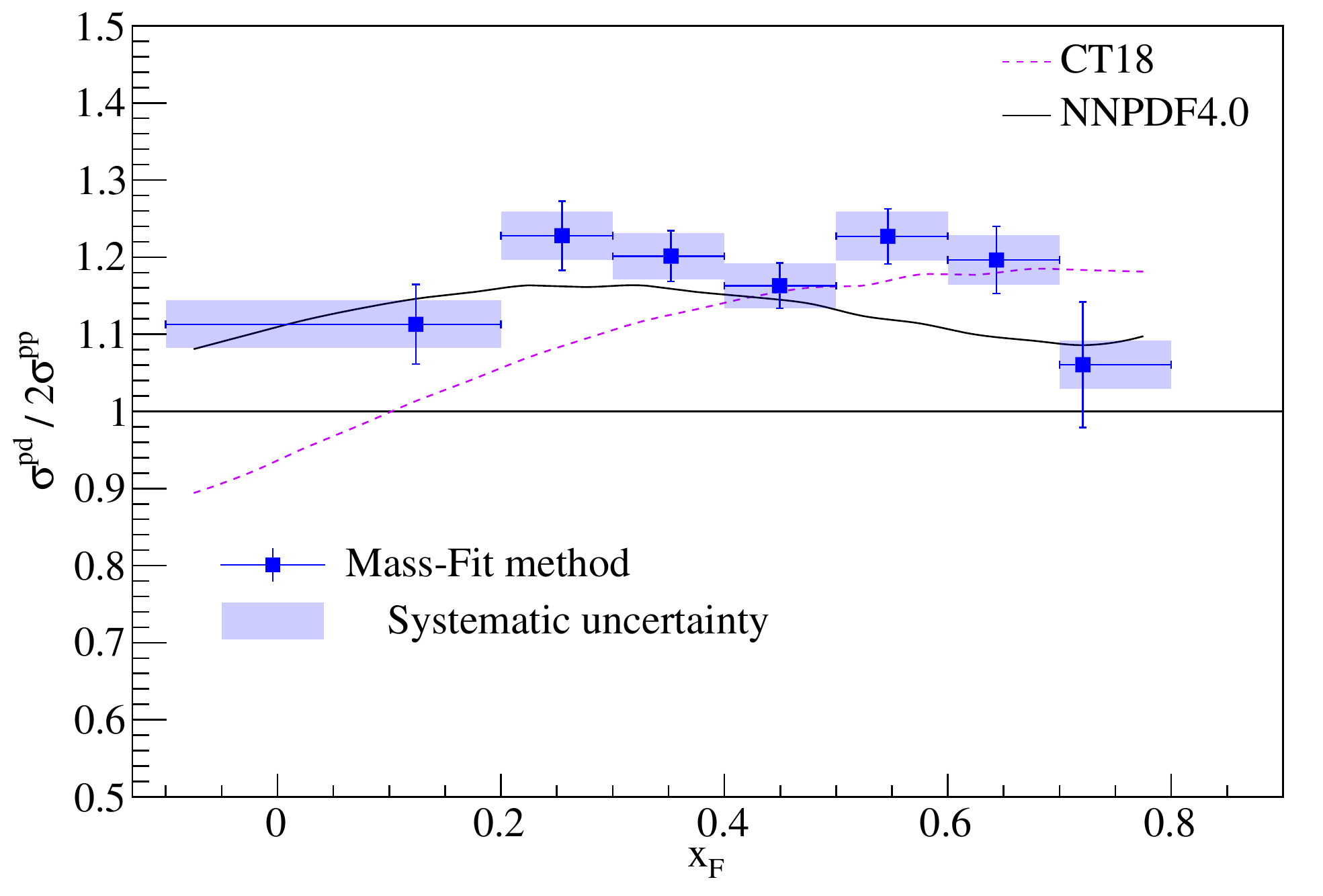}
    \caption{The cross section ratios versus $x_1$ (top) and $x_F$ (bottom).
    The curves represent calculations using two different proton parton distributions, the CT18 (red, dotted) and the NNPDF4.0 (black, solid), weighted by the SeaQuest acceptance.
    \label{fig:CSRII} }
\end{figure}

After the mass fit is performed to determine the normalization of the accidental background, a mass cut ($M > 4.5$ GeV) is applied to remove the $J/\psi$ and $\psi^\prime$ events. The remaining events can be projected onto various kinematic variables such as $x_2$, $x_1$, $x_F$, and $P_T$. Several steps are required before the yields of the Drell-Yan events are obtained. First, the non-Drell-Yan components, based on the normalizations determined from the fit of the mass spectrum, are subtracted from the data. Second, the effective luminosity for each liquid target, obtained from the total number of incident protons passing the BIM cut and the amount of $H_2$ or $D_2$ in each target, is used to normalize the integrated luminosity between the two targets.  Finally, the reconstruction inefficiency and the DAQ dead time, which have a small but non-negligible dependence on the target type, are taken into account.

\subsection{Cross Section Ratio Results}
The $\sigma^{pd}/2\sigma^{pp}$ Drell-Yan cross section ratios, $R$, obtained with the MF method, are shown as a function of $x_2$, $x_1$, $x_F$, and $P_T$ in Figs.~\ref{fig:CSR} and \ref{fig:CSRII}.  These values are given in Tabs.~\ref{tab:x2ratio}-\ref{tab:ptratio}. 
The size of the statistical uncertainty is different between the two methods while the dataset is common.
It is because, in the IE method, the statistics of one $x_2$ bin affects the cross section ratio of another bin through Eq.~\ref{eq:ComPol2}.
The statistics in the MF method is uncorrelated bin-to-bin.

The overall systematic uncertainties for the MF method are calculated as the quadrature sum of the various contributions. 
Major systematic uncertainties are the beam flux normalization, 
the correction for the measurement efficiency of dimuons,
and the dependence on Monte Carlo parameters.
Since the effects of these systematic uncertainties are highly correlated for the two targets, 
they are largely canceled out in the cross section ratio.
The systematic uncertainties of the two methods are almost uncorrelated.

\begin{table*}
  \caption{The cross section ratios for each $x_2$ bin obtained with the Intensity Extrapolation (IE) method and the Mass Fit (MF) method. 
  The first uncertainty is statistical, the second systematic. 
  The average values for dimuon kinematic variables are also shown.}
  \label{tab:x2ratio}
  \begin{center}
    \begin{tabular}{ccccccccc}
\hline  \hline
$x_2$ range & $\langle x_2\rangle$ & $\langle x_2\rangle$
& $\langle x_1\rangle$ & $\langle x_F\rangle$ & $\langle P_T\rangle$ & $\langle M_{\mu^+\mu^-}\rangle$ 
& $\sigma^{pd}/2\sigma^{pp}$ & $\sigma^{pd}/2\sigma^{pp}$ \\
            & IE & MF  & &  & (GeV) &  (GeV)  &  IE &  MF \\ \hline
0.130--0.160 & 0.147 & 0.148 & 0.689 & 0.605 & 0.643 & 4.712 & $1.211 \pm 0.052 \pm 0.053$ & $1.248 \pm 0.044 \pm 0.033$ \\
0.160--0.195 & 0.179 & 0.178 & 0.614 & 0.492 & 0.695 & 4.893 & $1.141 \pm 0.043 \pm 0.025$ & $1.154 \pm 0.028 \pm 0.029$ \\
0.195--0.240 & 0.216 & 0.216 & 0.561 & 0.396 & 0.701 & 5.152 & $1.196 \pm 0.042 \pm 0.044$ & $1.213 \pm 0.030 \pm 0.030$ \\
0.240--0.290 & 0.263 & 0.263 & 0.533 & 0.318 & 0.730 & 5.543 & $1.165 \pm 0.046 \pm 0.032$ & $1.184 \pm 0.038 \pm 0.030$ \\
0.290--0.350 & 0.315 & 0.315 & 0.512 & 0.240 & 0.752 & 5.955 & $1.193 \pm 0.050 \pm 0.034$ & $1.195 \pm 0.047 \pm 0.030$ \\
0.350--0.450 & 0.385 & 0.385 & 0.505 & 0.154 & 0.771 & 6.551 & $1.113 \pm 0.064 \pm 0.039$ & $1.131 \pm 0.076 \pm 0.032$ \\
\hline  \hline
    \end{tabular}
  \end{center}
\end{table*}

\begin{table*}
  \caption{The cross section ratio for each $x_1$ bin obtained with the Mass Fit (MF) method. 
  The first uncertainty is statistical, the second systematic. 
  The average values for dimuon kinematic variables are also shown.}
  \label{tab:x1ratio}
  \begin{center}
    \begin{tabular}{ccccccc}
\hline  \hline
$x_1$ range & $\langle x_1\rangle$
& $\langle x_2\rangle$ & $\langle x_F\rangle$ & $\langle P_T\rangle$ & $\langle M_{\mu^+\mu^-}\rangle$ 
& $\sigma^{pd}/2\sigma^{pp}$ \\
             &  &  &  & (GeV) &  (GeV)  &  MF\\ \hline
0.300--0.400 & 0.385 & 0.317 & 0.077 & 0.528 & 5.206 & $1.009 \pm 0.113 \pm 0.037$ \\
0.400--0.500 & 0.459 & 0.270 & 0.214 & 0.670 & 5.220 & $1.187 \pm 0.039 \pm 0.031$ \\
0.500--0.550 & 0.525 & 0.239 & 0.327 & 0.712 & 5.237 & $1.188 \pm 0.036 \pm 0.030$ \\
0.550--0.600 & 0.575 & 0.220 & 0.406 & 0.714 & 5.249 & $1.147 \pm 0.033 \pm 0.029$ \\
0.600--0.650 & 0.624 & 0.207 & 0.479 & 0.736 & 5.291 & $1.257 \pm 0.038 \pm 0.032$ \\
0.650--0.700 & 0.674 & 0.195 & 0.551 & 0.731 & 5.333 & $1.194 \pm 0.040 \pm 0.031$ \\
0.700--0.800 & 0.741 & 0.180 & 0.648 & 0.728 & 5.354 & $1.190 \pm 0.038 \pm 0.032$ \\
\hline  \hline
    \end{tabular}
  \end{center}
\end{table*}

\begin{table*}
  \caption{The cross section ratio for each $x_F$ bin obtained with the Mass Fit (MF) method. 
  The first uncertainty is statistical, the second systematic. 
  The average values for dimuon kinematic variables are also shown.}
  \label{tab:xFratio}
  \begin{center}
    \begin{tabular}{ccccccc}
\hline  \hline
$x_F$ range & $\langle x_F\rangle$
& $\langle x_1\rangle$ & $\langle x_2\rangle$ & $\langle P_T\rangle$ & $\langle M_{\mu^+\mu^-}\rangle$ 
& $\sigma^{pd}/2\sigma^{pp}$ \\
              &  &  &  & (GeV) &  (GeV)  & MF \\ \hline
-0.100--0.200 & 0.124 & 0.435 & 0.327 & 0.674 & 5.624 & $1.113 \pm 0.052 \pm 0.031$ \\
0.200--0.300 & 0.255 & 0.486 & 0.266 & 0.714 & 5.340 & $1.228 \pm 0.045 \pm 0.031$ \\
0.300--0.400 & 0.352 & 0.538 & 0.232 & 0.704 & 5.242 & $1.201 \pm 0.033 \pm 0.030$ \\
0.400--0.500 & 0.449 & 0.598 & 0.206 & 0.719 & 5.201 & $1.163 \pm 0.029 \pm 0.029$ \\
0.500--0.600 & 0.546 & 0.663 & 0.186 & 0.717 & 5.192 & $1.227 \pm 0.036 \pm 0.032$ \\
0.600--0.700 & 0.644 & 0.730 & 0.167 & 0.707 & 5.146 & $1.196 \pm 0.044 \pm 0.032$ \\
0.700--0.800 & 0.721 & 0.781 & 0.144 & 0.665 & 4.946 & $1.061 \pm 0.082 \pm 0.031$ \\
\hline  \hline
    \end{tabular}
  \end{center}
\end{table*}

\begin{table*}
  \caption{The cross section ratios for each $P_T$ bin obtained with the Intensity Extrapolation (IE) method and the Mass Fit (MF) method. 
  The first uncertainty is statistical, the second systematic. 
  The average values for dimuon kinematic variables are also shown.}
  \label{tab:ptratio}
  \begin{center}
    \begin{tabular}{ccccccccc}
\hline  \hline
$P_T$ range & $\langle P_T\rangle_{\textrm{IE}}$ & $\langle P_T\rangle_{\textrm{MF}}$
& $\langle x_1\rangle$ & $\langle x_2\rangle$ & $\langle x_F\rangle$ & $\langle M_{\mu^+\mu^-}\rangle$ 
& $\sigma^{pd}/2\sigma^{pp}$ & $\sigma^{pd}/2\sigma^{pp}$ \\
  (GeV)   & (GeV) & (GeV)  &  &  &  &  (GeV)  &  IE &  MF \\ \hline
0.000--0.300 & 0.198 & 0.197 & 0.569 & 0.222 & 0.397 & 5.225 & $1.137 \pm 0.049 \pm 0.061$ & $1.132 \pm 0.036 \pm 0.029$ \\
0.300--0.500 & 0.405 & 0.405 & 0.571 & 0.222 & 0.399 & 5.225 & $1.174 \pm 0.045 \pm 0.052$ & $1.164 \pm 0.031 \pm 0.029$ \\
0.500--0.700 & 0.599 & 0.599 & 0.572 & 0.224 & 0.399 & 5.240 & $1.209 \pm 0.046 \pm 0.038$ & $1.207 \pm 0.032 \pm 0.030$ \\
0.700--0.900 & 0.797 & 0.796 & 0.577 & 0.225 & 0.403 & 5.250 & $1.210 \pm 0.046 \pm 0.045$ & $1.249 \pm 0.037 \pm 0.031$ \\
0.900--1.200 & 1.035 & 1.033 & 0.580 & 0.230 & 0.402 & 5.300 & $1.130 \pm 0.043 \pm 0.037$ & $1.126 \pm 0.035 \pm 0.029$ \\
1.200--1.500 & 1.330 & 1.325 & 0.601 & 0.239 & 0.417 & 5.434 & $1.287 \pm 0.061 \pm 0.094$ & $1.312 \pm 0.080 \pm 0.037$ \\
1.500--1.800 & 1.625 & 1.620 & 0.606 & 0.250 & 0.411 & 5.520 & $1.087 \pm 0.078 \pm 0.099$ & $1.156 \pm 0.140 \pm 0.039$ \\
1.800--2.300 & 1.915 & 1.919 & 0.587 & 0.262 & 0.375 & 5.449 & $0.838 \pm 0.095 \pm 0.162$ & $0.898 \pm 0.183 \pm 0.045$ \\
\hline  \hline
    \end{tabular}
  \end{center}
\end{table*}

The Drell-Yan cross section ratios obtained with the two different analysis methods are in excellent agreement, as shown in Fig.~\ref{fig:CSR}.  The systematic uncertainty of the MF method is slightly smaller than that of the IE method. The good agreement between the two different analysis methods suggests that the results from either method can be used. Since the IE method was used in the recent paper~\cite{Dove21}, these cross section ratios will be used for further analysis to extract the values of $\bar d\left(x\right) / \bar u\left(x\right)$ and $\bar d\left(x\right) - \bar u\left(x\right)$, as described in Sec.~\ref{sec:extraction}.
  
To compare the cross section ratios with PDFs, the double differential Drell-Yan cross section, ${d\sigma} / {\left(dx_1dx_2\right)}$, is calculated at NLO using the DYNNLO code~\cite{Catani:2009sm,Catani:2007vq} for $p+p$ and $p+d$. To compare the calculation with the cross section ratios projected into the kinematic variables $x_1$, $x_2$, and $x_F$, the spectrometer acceptance as a function of $x_1$ and $x_2$ is needed, which are tabulated in Ref.~\cite{Dove21}.

% Discussion of CSR vs x2
The measured Drell-Yan cross section ratios versus $x_2$ are consistent with the results from E866 and calculations using the CT18~\cite{CT18} PDFs at low $x_2$, as shown in Fig.~\ref{fig:CSR}. Since the high statistics data at low $x_2$ from E866 were used to constrain the CT18 PDFs, the good agreement reflects the  consistency between E866 and SeaQuest results in the low $x_2$ region. 
In contrast, the data at higher $x_2$ are significantly higher than calculations using the CT18 PDFs. 
The CT18 PDFs, based on fits that include the E866 data, show a drop of the cross section ratios at high $x_2$. Such a drop of the cross section ratios at large $x_2$ is not observed in the SeaQuest experiment. 
The NNPDF4.0 PDFs~\cite{NNPDF40}, which include the recently published results~\cite{Dove21} from SeaQuest in the global fit, are naturally in much better agreement with the data.  

% Discussion of CSR vs pT
The cross section ratio versus $P_T$ in Fig.~\ref{fig:CSR} shows consistent results between the two methods. 
The drop at high $P_T$ is consistent with the E866 results~\cite{Towell:2001nh}, although this happens at lower $P_T$ in the SeaQuest data.  

% Discussion of CSR vs x1
The $x_1$ and $x_F$ dependence of the cross section ratios is shown in Fig.~\ref{fig:CSRII}. The data are compared with calculations with the CT18 and NNPDF4.0 PDFs. The data are overall higher than the calculations, which is consistent with the observation on the cross section ratios versus $x_2$.
Small-$x_1$ point corresponds to large-$x_2$ point, because of the spectrometer acceptance shown in Fig.~\ref{fig:acceptance_x12}.

% Discussion of CSR vs xF
%The $x_F$ dependence of the cross section ratios is shown in Fig.~\ref{fig:CSRII} with calculations with the CT18 and NNPDF4.0 PDFs. The data are overall higher than the calculations, as observed on the cross section ratios versus $x_1$.
%In the low $x_F$ region data are higher than calculations using either PDFs. 
%This region corresponds to the high $x_2$ region where the PDFs are influenced by the E866 data.  
%Much better agreement between the E906 data and the PDFs is observed at $x_F > 0.2$.

\section{Extraction of $\bar d\left(x\right) / \bar u\left(x\right)$ and $\bar d\left(x\right) - \bar u\left(x\right)$\label{sec:extraction}}

The $\bar d\left(x\right)/\bar u\left(x\right)$ ratio is extracted from the cross section ratio by an iterative method, as it has been done in Ref.~\cite{Dove21}.
It is primarily to demonstrate the significance of the cross section ratio results to the determination of the $\bar d\left(x\right)/\bar u\left(x\right)$ ratio.
Various simplifications are applied to the handling of PDFs as described below.
The extracted result is expected to be superseded by full global QCD fits.

An estimate for the $\bar d\left(x\right) / \bar u\left(x\right) $ ratio over the measured $x_2$ range is made and the cross section ratio $R$ is calculated with a chosen PDF set. This PDF set provides the parton distributions, except that the $\bar d\left(x\right) / \bar u\left(x\right)$  ratio, which is allowed to vary while keeping $\bar d\left(x\right) + \bar u\left(x\right)$ fixed at the value from the chosen PDF set. The $\bar d / \bar u$ ratio is initialized as 
\begin{equation}
    \left[\frac{\bar{d}}{\bar{u}}\right]\left(ix_2\right)= 2 R\left(ix_2\right)-1.
\end{equation}
Here $ix_2$ is the bin index for $x_2$.   The final result is insensitive to the initialization of the $\bar{d}/\bar{u}$ ratio.  
The $\bar{d}/\bar{u}$ ratio outside the measured $x_2$ range was assumed to be 1.0, and changed to 0.5 and 2.0 to estimate the systematic uncertainty of the assumption. There is an additional systematic effect arising from the uncertainties within the PDF fits themselves that is best addressed in the context of a global fit and not included here. The cross section ratio $R$ is then calculated using the NLO code as a function of $x_1$ and $x_2$, and summed over $x_1$ as
\begin{equation}
R_{\textrm{pred}}\left(ix_2\right) = 
\frac{\displaystyle   \sum_{ix_1} A_{ix_1,ix_2}\sigma^{pd}_{\text{NLO}}\left(ix_1,ix_2\right)}
     {\displaystyle 2 \sum_{ix_1} A_{ix_1,ix_2}\sigma^{pp}_{\text{NLO}}\left(ix_1,ix_2\right)},
\end{equation}
where $A_{ix_1,ix_2}$ is the acceptance for the $(x_1, x_2)$ bin.
The ratio $\bar{d}/\bar{u}$ is then adjusted according to the 
difference between the measured and predicted ratios, 
$\Delta_R\left(ix_2\right) = R_{\textrm{meas}}\left(ix_2\right)-R_{\textrm{pred}}\left(ix_2\right)$. 
This procedure is repeated until $\Delta_R\left(ix_2\right) < 10^{-3}$.

\begin{table}
  \caption{The $\bar d / \bar u$ and $\bar d - \bar u$ values for each $x$ bin.
  The first uncertainty is statistical, the second systematic.
  The asymmetry between the upper and lower uncertainty values reflects the non-linearity of the propagation of the symmetric cross section ratio uncertainties. 
  For $\bar{d}/\bar{u}$, $\left<M_{\mu^+\mu^-}\right>$ is the same as Tab.~\ref{tab:x2ratio}. 
  As explained in the text, $\bar{d}-\bar{u}$ is scaled to $Q^2 = 25.5$ GeV$^2$.
  }
  \label{tab:dbarratio}
  \begin{center}
    \begin{tabular}{c@{\hskip 10pt}c@{\hskip 10pt}c@{\hskip 10pt}c}
\hline
\hline%
% Note the extra room with the *[3 pt] that is used to space the super and subscripted errors
$x$ range & $\langle x \rangle$  
&$\bar{d}/\bar{u}$ & $ \bar{d} - \bar{u}$ \\ \hline 
\\*[-9 pt]
0.130--0.160 & 0.147 & $1.423 _{-0.089}^{+0.089} {}_{-0.103}^{+0.104}$ & $0.191 _{-0.035}^{+0.032} {}_{-0.040}^{+0.037}$ \\*[3 pt]
0.160--0.195 & 0.179 & $1.338 _{-0.085}^{+0.083} {}_{-0.065}^{+0.065}$ & $0.094 _{-0.021}^{+0.019} {}_{-0.016}^{+0.015}$ \\*[3 pt]
0.195--0.240 & 0.216 & $1.487 _{-0.092}^{+0.092} {}_{-0.110}^{+0.111}$ & $0.071 _{-0.011}^{+0.010} {}_{-0.013}^{+0.012}$ \\*[3 pt]
0.240--0.290 & 0.263 & $1.482 _{-0.113}^{+0.114} {}_{-0.097}^{+0.098}$ & $0.036 _{-0.007}^{+0.006} {}_{-0.006}^{+0.006}$ \\*[3 pt]
0.290--0.350 & 0.315 & $1.645 _{-0.140}^{+0.144} {}_{-0.121}^{+0.125}$ & $0.021 _{-0.004}^{+0.003} {}_{-0.003}^{+0.003}$ \\*[3 pt]
0.350--0.450 & 0.385 & $1.578 _{-0.203}^{+0.214} {}_{-0.148}^{+0.153}$ & $0.007 _{-0.002}^{+0.002} {}_{-0.001}^{+0.001}$ \\*[3 pt]
\hline
\hline
    \end{tabular}
  \end{center}
\end{table}

\begin{figure} \centering
    % note the use of \hfil and \hfill -- 2 vs 1 "l"s
    \includegraphics[width=\columnwidth]{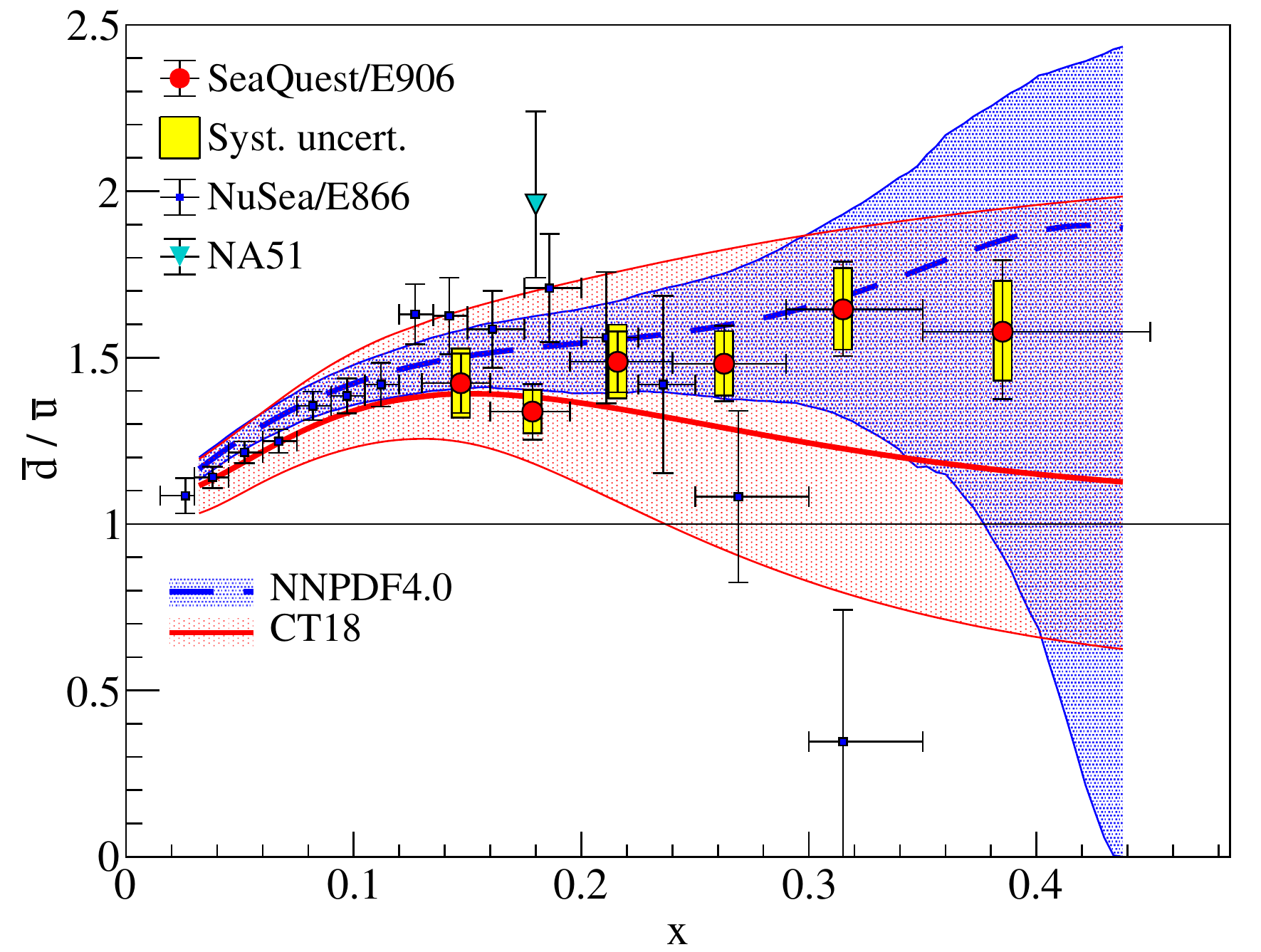} \\ %\hfil
    \includegraphics[width=\columnwidth]{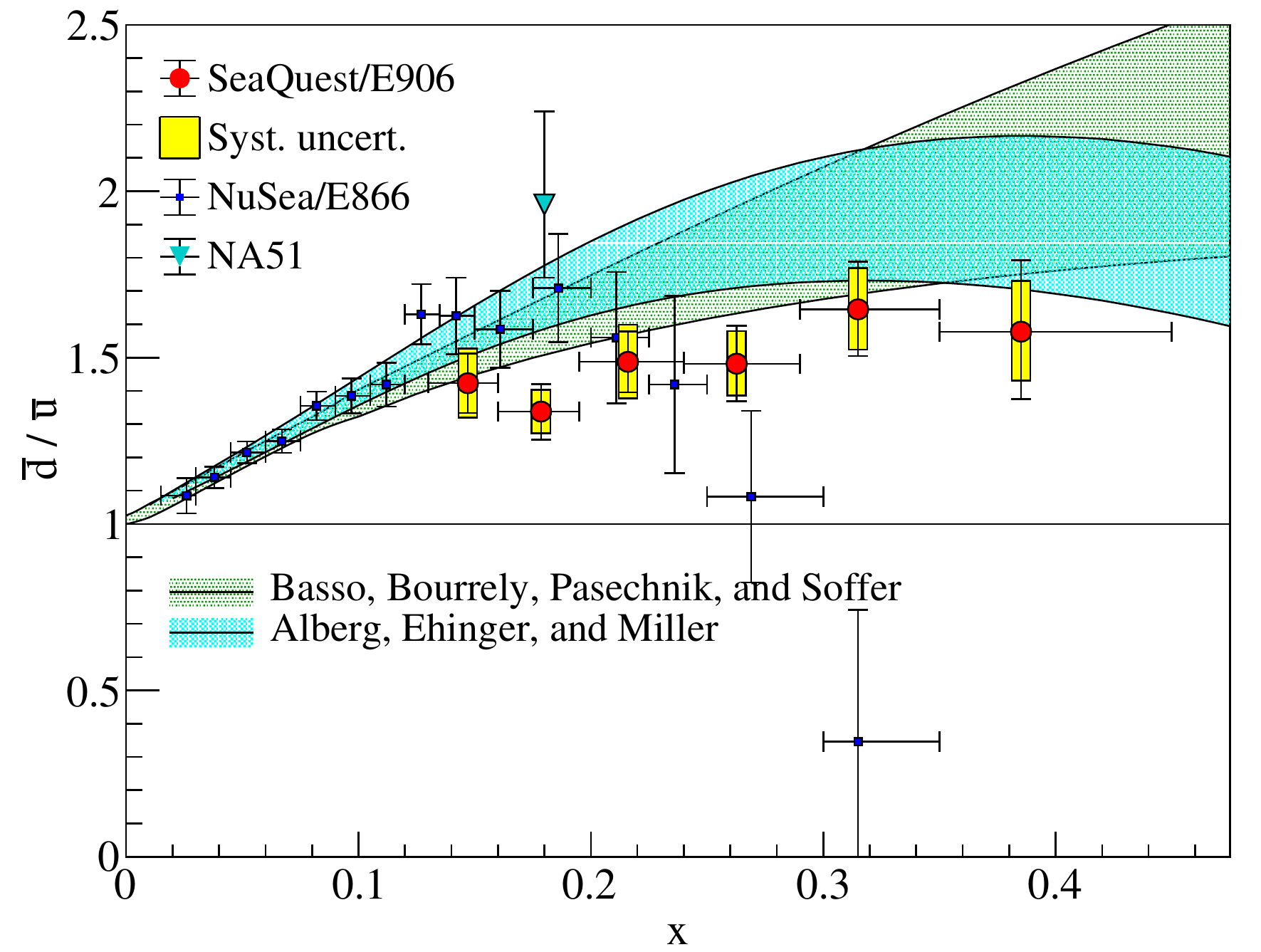}%\hfill
    \caption{The $\bar{d}/\bar{u}$ ratio versus $x$ based on NA51, E866, and SeaQuest data. The extracted ratios are compared with the CT18 and NNPDF4.0 PDFs (top), and with calculations in the meson-cloud model~\cite{Alberg, Alberg2} and statistical model~\cite{Basso} (bottom).\label{fig:dbarubar_xT}}
\end{figure}

The $\bar{d}(x) / \bar{u}(x)$ ratios obtained from the measured SeaQuest cross section ratios, using the CT18 PDFs as the basis for the extraction, are displayed in Fig.~\ref{fig:dbarubar_xT} and listed in Tab.~\ref{tab:dbarratio}. The ratios are consistent with the previous E866 data at low $x$.   At high $x$, the SeaQuest data show a slight continued rise in the ratio, and is in tension with the E866 data. The cause of the drop in the E866 data above $x = 0.2$ remains unexplained. The $\bar d(x)/\bar u(x)$ ratios are compared with the ratios with two PDFs, CT18 and NNPDF4.0, in Fig.~\ref{fig:dbarubar_xT}.  The SeaQuest data is in agreement with the meson-cloud~\cite{Alberg, Alberg2} and statistical models~\cite{Basso, Bourrely:2002vg, Bourrely:2005tk}.

\begin{table*}
  \caption{Values of $\int_{0.13}^{0.45} \left[\bar d\left(x\right) - \bar u\left(x\right)\right] dx$
  and $\int_{0.13}^{0.45}  x \left[\bar d\left(x\right) - \bar u\left(x\right)\right] dx$
  at $Q^2 = 25.5$ GeV$^2$ for two PDFs and the statistical model. 
The values deduced from SeaQuest are also listed. }
  \label{tab:tabdmu}
%  \begin{tabular}{c@{\hskip 15pt}|@{\hskip 15pt}c@{\hskip 15pt}|@{\hskip 15pt}cc@{\hskip 15pt}|@{\hskip 15pt}cc|}
  \begin{tabular}{c@{\hskip20pt}c@{\hskip 20pt}cc@{\hskip 20pt}cc}
\hline
\hline
 & & \multicolumn{2}{c}{PDFs} & \multicolumn{2}{c}{Models} \\
 & SeaQuest & CT18 & NNPDF4.0 & Stat. & Meson Cloud\\ \hline 
\\*[-9 pt]
$\int_{0.13}^{0.45} \left[\bar d\left(x\right) - \bar u\left(x\right)\right] dx$
& $0.0159 _{-0.0030}^{+0.0028} {}_{-0.0030}^{+0.0028}$ & $0.0129 ^{+0.0105}_{-0.0075}$ & $0.0208 ^{+0.0036}_{-0.0036}$ & $0.0186$ & $0.0180$ \\*[3 pt]
$\int_{0.13}^{0.45} x \left[\bar d\left(x\right) - \bar u\left(x\right)\right] dx$
& $0.00318 _{-0.00062}^{+0.00057} {}_{-0.00059}^{+0.00055}$ & $0.00241 ^{+0.00244}_{-0.00170}$ & $0.00414 ^{+0.00078}_{-0.00078}$ & $0.00386$ & $0.00361$ \\*[3 pt]
\hline
\hline
\end{tabular}
\end{table*}

From the $\bar d\left(x\right) / \bar u\left(x\right)$ ratios measured in SeaQuest, $\bar d\left(x\right) - \bar u\left(x\right)$ can be determined over the region $0.13 < x <0.45$. As a flavor non-singlet quantity, the integral, $\int_0^1\left[\bar d\left(x\right) - \bar u\left(x\right)\right]dx$, is $Q^2$ independent, even though the $x$-distribution may depend on $Q^2$. The $\bar d\left(x\right) - \bar u\left(x\right)$ values also provide a direct measure of the contribution from non-perturbative processes, since perturbative processes cannot cause a significant $\bar d$, $\bar u$ difference. These important properties of $\bar d\left(x\right) - \bar u\left(x\right)$ have been explored to extract the content of intrinsic light-quark sea of the proton~\cite{chang11,Brodsky,arXiv:2011.14971}. 
The flavor asymmetry in the proton's light quark sea has also been explored using lattice techniques within the framework of LaMET~\cite{Lin18,Cyprus21}. % \cite{Constantinou:2020hdm}.

\begin{figure}[tb]
\includegraphics[width=\columnwidth]{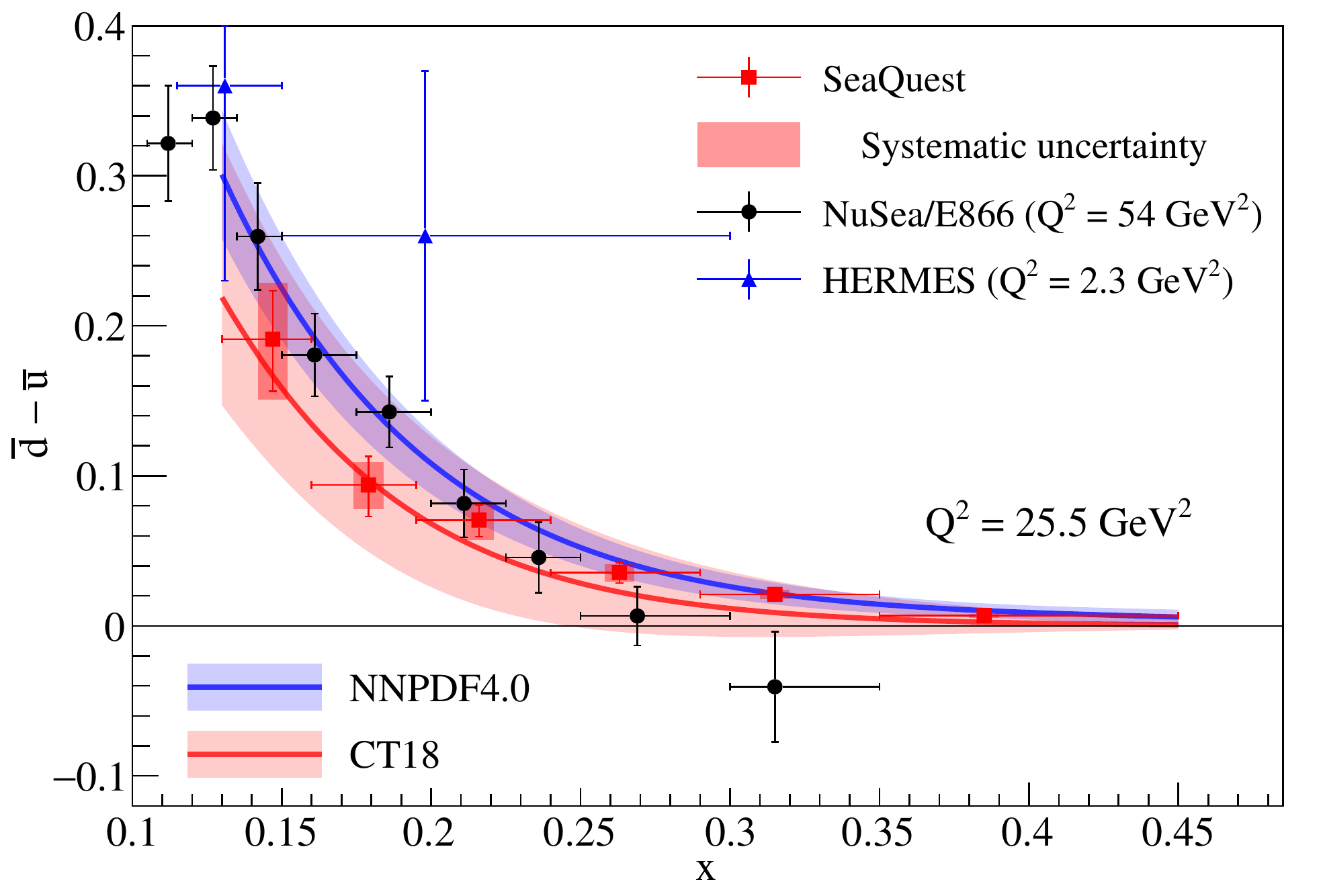} \\
\includegraphics[width=\columnwidth]{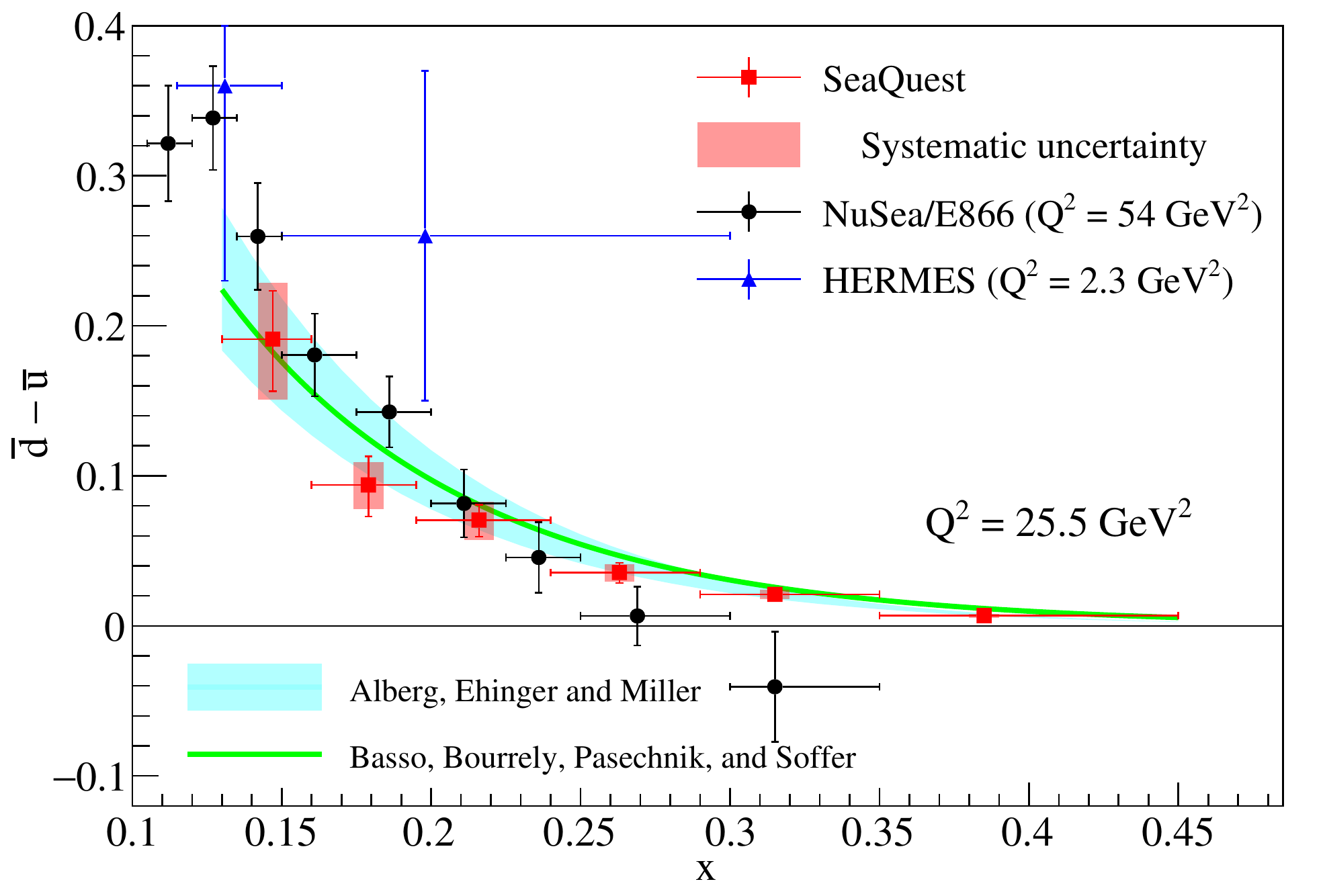}
\caption{The SeaQuest $\bar d\left(x\right) - \bar u\left(x\right)$ results compared with data from HERMES~\cite{Hermes} and E866~\cite{PhysRevLett.80.3715, Towell:2001nh} with calculations based on PDFs of CT18, NNPDF4.0 (top) and with the statistical models of Bourrely and Soffer~\cite{Basso, Bourrely:2002vg, Bourrely:2005tk}, and of Alberg, Ehinger and Miller~\cite{Alberg, Alberg2}.
\label{fig:dmu}}
\end{figure}

It is easiest to compare $\bar d\left(x\right) - \bar u\left(x\right)$ with models at a fixed $Q^2$. The $Q^2$ dependence of the data is obtained by scaling the values of $\bar d\left(x\right)/\bar u\left(x\right)$ obtained at the mean values of $Q^2$ for each $\langle x_2\rangle$ bins to a fixed value of $Q^2=25.5$ GeV$^2$ by following the $Q^2$-dependence of $\bar d\left(x\right) / \bar u\left(x\right)$ of CT18 PDF as an input. The $Q^2$ dependencies are below 2\% for all $x_2$ bins.  The values of $\bar d\left(x\right) + \bar u\left(x\right)$ at $Q^2= 25.5$ GeV$^2$ from CT18 are used to convert the values of $\bar d\left(x\right) / \bar u\left(x\right)$ into $\bar d\left(x\right) - \bar u\left(x\right)$. These values are insensitive to the specific PDFs used in this procedure. The values extracted from the SeaQuest data at $Q^2=25.5$ GeV$^2$ over the region $0.13 < x < 0.45$ are shown in Fig.~\ref{fig:dmu} and given in Tab.~\ref{tab:dbarratio}.

%Figure~\ref{fig:dmu} and Table~\ref{tab:dbarratio} show the values of $\bar d\left(x\right) - \bar u\left(x\right)$ extracted from SeaQuest at $Q^2=25.5$ GeV$^2$ over the region $0.13 < x < 0.45$. We first \bc {scale} the values of $\bar d\left(x\right)/\bar u\left(x\right)$, obtained at different mean values of $Q^2$ for different $x_2$ shown in Fig.~\ref{fig:dbarubar_xT}, to a fixed value of $Q^2=25.5$ GeV$^2$ by utilzing the $Q^2$-depedence of $\bar d\left(x\right) / \bar u\left(x\right)$ of CT18 PDF as an input. We find that the $Q^2$ dependences are below 2\% for all $x_2$ bins.  We then use the values of $\bar d\left(x\right) + \bar u\left(x\right)$ at $Q^2= 25$ GeV$^2$ from CT18 to convert the values of $\bar d\left(x\right) / \bar u\left(x\right)$ into $\bar d\left(x\right) - \bar u\left(x\right)$. We find that the values of $\bar d\left(x\right) - \bar u\left(x\right)$ are insensitive to the specific PDFs used in this procedure. 

The $\bar d\left(x\right) - \bar u\left(x\right)$ values derived from the SeaQuest data are compared with data from the HERMES~\cite{Hermes} and E866~\cite{PhysRevLett.80.3715, Towell:2001nh} experiments, and with calculations from the meson cloud and the statistical models in Fig.~\ref{fig:dmu}. The quantity $\bar d(x) - \bar u(x)$ is insensitive to the flavor-symmetric components of the sea-quark distributions in these models~\cite{Peng:1998pa}, unlike $\bar d\left(x\right) / \bar u\left(x\right)$. The SeaQuest data is in good agreement with both the statistical model and the meson cloud model.

The SeaQuest data on $\bar d\left(x\right) - \bar u\left(x\right)$ can also provide a determination of two integrals: $\int_{0.13}^{0.45} \left[\bar d\left(x\right) - \bar u\left(x\right)\right] dx$, which characterizes the integrated sea-quark flavor asymmetry, and $\int_{0.13}^{0.45} x \left[\bar d\left(x\right) - \bar u\left(x\right)\right] dx$, which represents the difference of the momentum fractions carried by $\bar d$ and $\bar u$ sea quarks. These values are listed in Tab.~\ref{tab:tabdmu}. The SeaQuest results are also compared with the calculations from two different PDFs, as well as from the meson cloud and statistical models. The NMC experiment obtained $\int_0^1 \left[\bar{d}\left(x\right)-\bar{u}\left(x\right)\right] dx = 0.147 \pm 0.039$ based on their measurements over the region $0.004 < x < 0.8$ and extrapolations outside of their measured $x$ region~\cite{NMC91,NMC2}. The E866 experiment found $\int_0^1 \left[\bar{d}\left(x\right)-\bar{u}\left(x\right)\right] dx = 0.118 \pm 0.012$ based on the measurement over the region $0.015 < x < 0.35$ and an extrapolations~\cite{Towell:2001nh}. Compared to these values, $\int_{0.13}^{0.45}\left[\bar d(x) - \bar u(x)\right]dx = 0.0159$ from SeaQuest is small, mainly because the PDF's themselves are small in this $x$ range. 

\section{Conclusions\label{sec:conclusions}}

Results from the SeaQuest experiment on the Drell-Yan $\left(p+d\right)/2\left(p+p\right)$ cross section ratios in the large $x$ region up to $x=0.45$ are reported. The SeaQuest experiment was designed to improve the accuracy of previous Drell-Yan experiments. Using an intense 120 GeV proton beam on identical liquid hydrogen and deuterium targets and a dimuon spectrometer constructed for this experiment, cross section ratios covering the range $0.13 < x < 0.45$ have been measured. Two different approaches were used to determine the cross section ratios. The results as functions of $x_2$ and $P_T$ were found consistent, indicating the robustness of the SeaQuest measurement. The cross section ratios as functions of $x_1$ and $x_F$ were newly obtained. The ratios of $\bar d\left(x\right) / \bar u\left(x\right)$ were  extracted from the measured cross section ratios and compared with those of the E866 experiment. The SeaQuest result shows that the $\bar d\left(x\right) / \bar u\left(x\right)$ ratios are greater than unity for the entire measured $x$ range. While the SeaQuest result is in qualitative agreement with that of E866, some tension in the $\bar d\left(x\right) / \bar u\left(x\right)$ ratios at the largest $x$ remains. The origin of this apparent difference is not understood. Thus, future independent investigations of the flavor structure of nucleon sea may be warranted.

Several recent global analyses~\cite{NNPDF40,JAM,10.21468/SciPostPhysProc.8.005,10.21468/SciPostPhysProc.8.177} have included the new SeaQuest results~\cite{Dove21} in addition to the recent $W$-boson production data from the STAR collaboration~\cite{STAR}. These new proton PDFs have shown that the SeaQuest data significantly reduce the uncertainties of $\bar d / \bar u$ at large $x$. 
These new results of $\bar d\left(x\right) / \bar u\left(x\right)$ support theoretical models, including the meson-cloud and statistical models, which predict that these ratios continue to rise as $x$ increases. The SeaQuest data will place new stringent constraints on future efforts to extract proton parton distributions. 

The SeaQuest measurement of the ratios of Drell-Yan cross sections,  $\left(p+d\right)/2\left(p+p\right)$, also provides an extraction of the flavor non-singlet quantity $\bar d\left(x\right) - \bar u\left(x\right)$ over the region $0.13 < x < 0.45$. This quantity is independent of the perturbative QCD contributions to PDFs and is a valuable probe for non-perturbative QCD origins of the proton sea. 

As demonstrated by the HERMES experiment, the sea-quark flavor asymmetry can also be probed using the Semi-Inclusive DIS reaction~\cite{Hermes}. With the advent of the Electron Ion Collider, a measurement of $\bar d\left(x\right) - \bar u\left(x\right)$ via Semi-Inclusive DIS in the small $x$ region, where abundant sea quarks reside, becomes feasible. The measurement of the Gottfried sum can also be extended to cover the smaller $x$ region. An extension of the SeaQuest experiment, SpinQuest~\cite{SpinQuestWeb}, will search for possible flavor asymmetries in the Sivers functions using transversely polarized targets.

\section*{Acknowledgements\label{sec:ack}}

We thank G.~T.~Garvey for contributions to the early stages of this experiment.  We also thank the Fermilab Accelerator Division and Particle Physics Division for their support of this experiment. This work was performed by the SeaQuest Collaboration, whose work was supported in part by 
the US Department of Energy under grants 
DE-AC02-06CH11357, % confirmed by Paul Argonne
%DE-FG02-07ER41528, DE-SC0006963; 
the US National Science Foundation under grants 
PHY 1812340, % confirmed by Jen-Chieh
PHY 2012926, % confirmed by Christine
%PHY 1812377, % confirmed by Roy if we list him as CalTech
PHY 1807338, % confirmed by Wolfgang
PHY 2110229, %confirmed by Wolfgang
PHY 2111046, %UIUC, confirmed by Hugo
PHY 1506489; % confirmed by Ed Kinney
% Old NSF PHY 0969239, PHY 1306126, PHY 1452636, PHY 1505458, PHY 1614456; 
National Science and Technology Council of Taiwan (R.O.C.); %confirmed by Wen-Chen
the DP\&A and ORED at Mississippi State University; %confirmed by Lamiaa
the JSPS (Japan) KAKENHI through grant numbers 21244028, 25247037, 25800133, 20K04000, 22H01244; % confirmed by Kenichi
the Bilateral Programs Joint Research Projects of JSPS; % confirmed by Kenichi
the Tokyo Tech Global COE Program, Japan; % confirmed by Kenichi
the Yamada Science Foundation of Japan; % confirmed by Kenichi
and the Ministry of Science and Technology (MOST) of Taiwan.
Fermilab is managed by Fermi Research Alliance, LLC (FRA), acting under Contract No. DE-AC02-07CH11359.

\bibliographystyle{apsrev4-1}
\bibliography{Drell-Yan}

\end{document}